\documentclass[11pt]{article}
\usepackage{amssymb}
\usepackage{amsmath}
\usepackage{amscd}
\usepackage{latexsym}
\usepackage{ifthen}
\usepackage[xdvi]{graphics}

\catcode`@=11
\renewcommand{\section}{\@startsection{section}{1}{0pt}{\medskipamount}
{\medskipamount}{\large\bf}}
\numberwithin{equation}{section}
\catcode`@=12

\def\a{\alpha}
\def\b{\beta}
\def\g{\gamma}
\def\de{\delta}
\def\eps{\epsilon}
\def\ve{\varepsilon}
\def\z{\zeta}

\def\th{\theta}
\def\l{\lambda}

\def\r{\rho}
\def\s{\sigma}

\def\ps{\psi}
\def\o{\omega}
\def\t{\tau}
\def\P{\Phi}

\newcommand{\U}{\Omega} 
\newcommand{\C}{\mathbb C}
\newcommand{\R}{\mathbb R}

\newcommand{\N}{\mathbb N}
\newcommand{\Acal}{{\R_{\th}^{2,2}}}
\newcommand{\Hcal}{{\cal H}}

\newcommand{\Zcal}{{\cal Z}}
\newcommand{\Jcal}{{\cal J}}

\def\>{\rangle}
\def\<{\langle}
\def\N2{$N{=}2$}
\def\pa{\partial}

\def\tr{{\rm tr}}

\newcommand{\ad}{{\dot{\alpha}}}
\newcommand{\bd}{{\dot{\beta}}}

\newcommand{\eb}{{\bar{\eta}}}
\newcommand{\lb}{\bar{\l}}
\newcommand{\mb}{\bar{\mu}}

\newcommand{\Zb}{\bar{Z}}

\newcommand{\ib}{{\bar{\imath}}}
\newcommand{\jb}{{\bar{\jmath}}}
\newcommand{\zb}{\bar{z}}
\newcommand{\wb}{\overline{w}}
\newcommand{\ov}[1]{\overline{#1}}
\newcommand{\wt}{\widetilde{w}}
\newcommand{\tht}{\widetilde{\th}}

\newcommand{\Tt}{\widetilde{T}}

\newcommand{\Gt}{\widetilde{G}}

\newcommand{\ut}{{\tilde{u}}}
\newcommand{\vt}{{\tilde{v}}}
\newcommand{\wh}{\widehat{w}}
\newcommand{\Fh}{F}  
\newcommand{\wbh}{\widehat{\wb}}

\newcommand{\ic}{\text{i}}

\newcommand{\Gp}{\ifthenelse{\boolean{mmode}}{{G^+}}{\mbox{$G^+\:$}}}
\newcommand{\Gtp}{\ifthenelse{\boolean{mmode}}{\mbox{$\Gt^+$}}{\mbox{$\Gt^+\:$}}}
\newcommand{\Gm}{\ifthenelse{\boolean{mmode}}{{G^-}}{\mbox{$G^-\:$}}}
\newcommand{\Gtm}{\ifthenelse{\boolean{mmode}}{\mbox{$\Gt^-$}}{\mbox{$\Gt^-\:$}}}

\begin{document}
\begin{titlepage}
\setcounter{page}{0}
\begin{flushright}
{\tt hep-th/0211263}\\
ITP--UH--23/02\\
\end{flushright}

\vskip 2.0cm

\begin{center}

{\Large\bf  Noncommutative Extended Waves and Soliton-like\\[2mm]
            Configurations in N=2 String Theory}

\vspace{14mm}

{\large Matthias Ihl \ and \ Sebastian Uhlmann}
\\[5mm]
{\em Institut f\"ur Theoretische Physik  \\
Universit\"at Hannover \\
Appelstra\ss{}e 2, 30167 Hannover, Germany }\\[2mm]
{E-mail: {\tt msihl, uhlmann@itp.uni-hannover.de}}

\end{center}

\vspace{2cm}

\noindent
{\sc Abstract:} The Seiberg-Witten limit of fermionic N=2 string
theory with nonvanishing $B$-field is governed by noncommutative
self-dual Yang-Mills theory (ncSDYM) in 2+2 dimensions. Conversely,
the self-duality equations are contained in the equation of motion
of N=2 string field theory in a $B$-field background. Therefore
finding solutions to noncommutative self-dual
Yang-Mills theory on $\R^{2,2}$ might help to improve our
understanding of nonperturbative properties of string (field) theory.
In this paper, we construct nonlinear soliton-like and multi-plane wave
solutions of the \mbox{ncSDYM} equations corresponding to certain D-brane
configurations by employing a solution generating technique, an
extension of the so-called dressing approach.
The underlying Lax pair is discussed in two different gauges, the
unitary and the hermitean gauge. Several examples and applications for
both situations are considered, including abelian solutions
constructed from GMS-like projectors, noncommutative $U(2)$ soliton-like
configurations and interacting plane waves. We display a correspondence
to earlier work on string field theory and argue that the solutions
found here can serve as a guideline in the search for nonperturbative
solutions of nonpolynomial string field theory.\\[.7cm]

\vfill

\textwidth 6.5truein

\end{titlepage}


\noindent
\tableofcontents
\newpage

\section{Introduction}\label{sec:intro}
\noindent
The study of noncommutative field theory has become an important
subject in modern theoretical physics. Even though the idea of
noncommuting coordinates is a rather old one~\cite{Snyder:1946qz},
research in this direction has been boosted only after the discovery
that noncommutativity naturally emerges in string theory with a $B$-field
background in a certain zero slope ($\a'\to 0$) limit~%
\cite{Douglas:1997fm, Schomerus:1999ug, Seiberg:1999vs}. Of special
interest is the study of nonperturbative objects in its low energy
field theory limits, like solitons or instantons, which may be
interpreted as D-branes in the context of string theory (for a review
see~\cite{Nekrasov:2000ih, Harvey:2001yn, Douglas:2001ba, Konechny:2001wz}).
The goal is to gain some insight into the nonperturbative sector of
these theories.

The discovery of Ooguri and Vafa that open N=2 string theory at tree
level can be identified with self-dual Yang-Mills theory~\cite{Ooguri:1990ww}
sparked new interest in the study of this area. That noncommutative
self-dual Yang-Mills (ncSDYM) appears as the effective field theory
describing the open critical N=2 string in 2+2 dimensions with
nonvanishing $B$-field was shown later in~\cite{Lechtenfeld:2000nm,
Lechtenfeld:2001uq}. Furthermore, (commutative) self-dual Yang-Mills
has been conjectured to be a universal integrable model (\cite{Ward:gz},
see also \cite{Ivanova:rc} and references therein), meaning that
all (or at least most) of the integrable equations in $d<4$ can be
obtained from the self-dual Yang-Mills equations. Therefore it is
worthwhile to study the noncommutative generalization of this theory
and, more specifically, plane wave and soliton-like solutions to the
ncSDYM equations.

A lot of work concerning noncommutative solitons has been carried out
during the past years (see e.g.~\cite{Gopakumar:2000zd} --
\cite{Jatkar:2000ei}). In particular, noncommutative solitons and plane
waves in an integrable $U(N)$ sigma model in 2+1 dimensions were
discussed in~\cite{Lechtenfeld:2001aw} -- \cite{Wolf:2002jw}. In
the present work we will show that all the cases studied in~%
\cite{Lechtenfeld:2001aw} -- \cite{Wolf:2002jw} can be obtained from
ncSDYM theory by dimensional reduction, i.e., by demanding that the
solutions do not depend on one of the time directions (see section~%
\ref{sec:dimredux}). The self-duality equations will be regarded as
the compatibility conditions of two linear equations (Lax pair).
Solutions $\psi$ to residue equations of the latter can then be used
to find solutions to the self-duality equations. By employing a solution
generating technique, namely a noncommutative extension of the so-called
{\sl dressing approach}~\cite{Zakharov:pp, Forgacs:1983gr, Ward:ie},
we are able to compute the aforementioned auxiliary field $\psi$.
Starting from a simple first order pole ansatz for $\psi$, one can
easily construct higher order pole solutions corresponding to multi-soliton
configurations. Recently it has been shown that a variant of the dressing
method can be used to construct exact solutions of Berkovits' string
field theory~\cite{Lechtenfeld:2002cu,Kling:2002ht}. We will elucidate
the relation between the field theoretic solutions and possible solutions
in string field theory.

The paper is organized as follows: Section~\ref{sec:motivation} contains
a review of some results from N=2 string theory with nonvanishing $B$-field,
motivating the program carried out in this paper from a string theory
point of view and placing it in this context. In section~\ref{sec:sdym}
we introduce our notation and conventions as well as the Moyal-Weyl map as
a useful tool for later computations. After that, the dressing approach
will be discussed in section~\ref{sec:dress}; a clarification of the
relation between the field theory approach given herein and its string
field theoretic variant introduced in~\cite{Lechtenfeld:2002cu} is added
at the end of section~\ref{sec:dress}. We present various calculations
and examples of solutions in this framework in sections~%
\ref{sec:nointeract} and~\ref{sec:scattering}. The discussion of some
mathematical preliminaries like twistor spaces and the moduli space of
complex structures on $\R^{2,2}$ is relegated into appendix~%
\ref{sec:twistor}. An example of an abelian pseudo-instanton which is
somewhat detached from the rest of the paper will be discussed in
appendix~\ref{sec:energy}.

\section{Noncommutativity from string theory}\label{sec:motivation}
\noindent
{\bf N=0 and N=1 string theories.}
It is well known for N=0 and N=1 string theories that turning on a
$B$-field in the presence of D-branes modifies the dynamics of open
strings~\cite{Seiberg:1999vs}. It alters the ordinary Neumann boundary
conditions along the brane, which results in a deformation of the
space-time metric $G_{\mu\nu}$ seen by open strings. Another
consequence is the emergence of space-time noncommutativity in the
world-volume of the brane~\cite{Schomerus:1999ug},
\begin{equation}\label{eq:etc}
  [X^\mu(\tau),X^\nu(\tau)]= \ic \th^{\mu \nu}.
\end{equation}
This noncommutativity pertains to the low energy field theory capturing
the dynamics of open strings on the brane. In this discussion,
$G^{\mu\nu}$ and $\th^{\mu\nu}$ can be extracted from the closed
string metric $g_{\mu\nu}$ and the Kalb-Ramond field $B_{\mu\nu}$
as the symmetric and antisymmetric part of
\begin{equation}
  [(g+ 2\pi\a' B)^{-1}]^{\mu \nu} = G^{\mu\nu} +
    \frac{1}{2\pi\a'} \th^{\mu\nu}. \label{eq:GBdef}
\end{equation}
In the Seiberg-Witten limit~\cite{Seiberg:1999vs}
\begin{equation} \label{eq:SWl}
  \a' \to 0, \quad \text{keeping} \; G^{\mu\nu}\text{ and }\th^{\mu
    \nu} \; \text{fixed},
\end{equation}
open string theory reduces to noncommutative Yang-Mills.\footnote{%
Alternatively, one can keep $\a'$ and $g_{\mu\nu}$ fixed and take
$B\to\infty$. This formulation will be useful for the string field
theory discussion in section~\ref{sec:SFT}.} The effective open
string coupling $G_s$, which is related to the closed string coupling
$g_s$ via $G_s = g_s[(\text{det}G/ \text{det}(g+ 2 \pi \alpha'
B)]^{-1/2}$, in this limit reduces to
\begin{equation}
  G_s \stackrel{\alpha' \rightarrow 0}{\longrightarrow}
    \frac{g_{\text{YM}}^2}{2 \pi}.
\end{equation}
Note that bulk effects (due to closed string modes) only decouple from
the open string modes if we take the Seiberg-Witten limit.

It is a standard result (cf.~\cite{Taylor:1997dy} and references
therein) that soliton solutions in Yang-Mills \mbox{(-Higgs)}
theory can be interpreted as lower-dimensional D-brane configurations.
These induce an ``electric'' field $F_{\mu\nu}$ on the brane, thus the
$B$-field in the above formulas has to be replaced by the gauge invariant
quantity ${\cal F}_{\mu\nu}:= B_{\mu\nu}+F_{\mu\nu}$. The noncommutativity
parameter $\th^{\mu\nu}$ will in general be a function of
${\cal F}_{\mu\nu}$. Note that, in this paper, the back reaction of a
nonvanishing gauge field configuration on the open string parameters
will be neglected.

\noindent
{\bf N=2 string theory.}
In the case of (critical) N=2 fermionic string theory in 2+2 dimensions,
an analysis of $B$-field effects was carried out in~%
\cite{Lechtenfeld:2000nm}. In the following, we shall briefly delineate
the results of this paper. In critical N=2 string theory with
nonvanishing K\"ahler two-form field $B=(B_{\mu\nu})$, the dynamics of
fields on $N$ coincident space-time filling D-branes\footnote{Due
to the absence of R-R forms in the closed string spectrum of N=2
string theory, D-branes are simply defined in parallel to bosonic string
theory as submanifolds on which open strings can end.} in the
Seiberg-Witten limit is governed by $U(N)$ ncSDYM in the Leznov gauge.%
\footnote{N=2 string theory with $N$ coincident D2-branes yields an
integrable modified $U(N)$ sigma model on noncommutative $\R^{2,1}$~%
\cite{Lechtenfeld:2001uq} (generalizing the commutative case considered
by Ward~\cite{Ward:ie}).}
As a nontrivial check, the authors of~\cite{Lechtenfeld:2000nm} showed
the vanishing of the noncommutative field-theory four-point amplitude
at tree level. This is in accordance with the expectation from N=2
string theory, which features trivial $n$-point tree-level scattering
amplitudes for $n>3$, due to a certain kinematical identity in 2+2
dimensions. In this context it is worthwhile to emphasize two points:
The failure of the Moyal-Weyl commutator to close in $su(N)$ necessitates
the enlargement of the gauge group from $SU(N)$ to $U(N)$~%
\cite{Matsubara:2000gr}. Furthermore, to obtain ncSDYM in the Yang
gauge~\cite{Yang:1977zf}, which will mostly be used in this paper, one
has to consider N=2 string theory restricted to the zero world-sheet
instanton sector.

After this brief string theoretic overture, let us now turn to ncSDYM,
whose nonperturbative solutions shall concern us for the rest of this
paper.

\section{Noncommutative self-dual Yang-Mills on $\R^{2,2}$}\label{sec:sdym}

\subsection{Notation and conventions}
\noindent
In this paper we will consider solutions to the self-duality equations
for the noncommutative version of $U(N)$ Yang-Mills theory on the
space $\R^{2,2}$. We choose coordinates $(x^{\mu})=(x,y,\tilde{t},-t)$
such that the metric will take the form $(\eta_{\mu\nu})=\text{diag}
(+1,+1,-1,-1)$.\footnote{All conventions are chosen to match those of~%
\cite{Lechtenfeld:2001aw, Lechtenfeld:2002cu}. The choice $x^4=-t$ is
motivated by the fact that the hyperplane $\tilde{t}=0$ then has the
same orientation as in earlier work on self-dual Yang-Mills theory
dimensionally reduced to this hyperplane.}

\noindent
{\bf Coordinates.}
The signature $(++-\,-)$ allows for two different choices of isotropic
(light-like) coordinates (see appendix~\ref{sec:twistor}). The set of
real isotropic coordinates (suitable for the discussion of the unitary
gauge, see section~\ref{sec:unitary}) is
\begin{subequations}
\begin{eqnarray}
  u:={\frac{1}{2}}(t+y),\quad  & & v:={\frac{1}{2}}(t-y),\\
  \ut:={\frac{1}{2}}({\tilde{t}}+x),\quad  & &
    \vt:={\frac{1}{2}}({\tilde{t}}-x),
\end{eqnarray}
\end{subequations}
giving rise to
\begin{subequations}\label{eq:realder}
\begin{eqnarray}
  \pa_u=\pa_t + \pa_y,\quad  & & \pa_v= \pa_t - \pa_y, \\
  \pa_\ut = \pa_{\tilde{t}} + \pa_x,\quad & &
    \pa_\vt= \pa_{\tilde{t}} - \pa_x.
\end{eqnarray}
\end{subequations}
For the discussion of the hermitean gauge (section~\ref{sec:hermitean}),
the other choice of isotropic coordinates, namely complex ones,
\begin{subequations}\label{eq:cplx}
\begin{eqnarray}
  z^1:= x + \ic y, \quad & & \zb^1= x - \ic y, \\
  z^2:= \tilde{t} - \ic t, \quad & & \zb^2 = \tilde{t} + \ic t,
\end{eqnarray}
\end{subequations}
turns out to be useful. These yield the following partial derivatives
\begin{subequations}
\begin{eqnarray}
  \pa_{z^1}= \frac{1}{2} (\pa_x - \ic \pa_y),\quad  & &
    \pa_{\zb^1}= \frac{1}{2} (\pa_x + \ic \pa_y), \\
  \pa_{z^2}= \frac{1}{2}(\pa_{\tilde{t}} + \ic \pa_t),\quad & &
    \pa_{\zb^2}= \frac{1}{2}(\pa_{\tilde{t}} - \ic \pa_t).
\end{eqnarray}
\end{subequations}

\noindent
{\bf Star product.} The multiplication law used to multiply functions
is the standard Moyal-Weyl star product given by
\begin{equation}\label{eq:mws}
  (f \star g)(x):= \text{e}^{\left[{\frac{\ic}{2}}
    (\th^{\mu\nu}\frac{\pa}{\pa x^\mu} \frac{\pa}{\pa y^\nu})
    \right]} f(x)g(y) \Big{|}_{y=x}.
\end{equation}
The noncommutativity of the coordinates is encoded in the usual
structure of the commutator\footnote{$[x^\mu \stackrel{\star}{,}
x^\nu] := x^\mu \star x^\nu -  x^\nu \star x^\mu$.}
\begin{equation}
  [x^\mu{\stackrel{\star}{,}}x^\nu] = \ic \th^{\mu\nu}.
\end{equation}
As a constant antisymmetric matrix, $\th^{\mu\nu}$ is taken to be
\begin{equation}
  (\th^{\mu\nu}):=\left( \begin{array}{cccc}
    0 & \th^{12} & 0 & 0 \\
    \th^{21} & 0 & 0 & 0 \\
    0 & 0 & 0 & \th^{34}\\
    0 & 0 & \th^{43} & 0 \end{array} \right),
\end{equation}
where
$\th^{12}=-\th^{21}=:\th$ and $\th^{34}=-\th^{43}=:\tht$. Without loss
of generality we assume $\th>0$, and choose $\tht\geq 0$, which, in the
case $\th=\tht$, corresponds to self-dual $\th^{\mu \nu}$.%
\footnote{In the self-dual case, $\th_{\mu \nu} = \frac{1}{2}
\epsilon_{\mu \nu \rho \sigma} \th^{\rho \sigma}$, where
$\epsilon_{1234}:=1$.}
Note that we are dealing with two time directions which mutually
{\sl do not} commute, but that the commutator of one temporal and one
spatial coordinate still vanishes.

\noindent
{\bf Yang-Mills theory.}
The action of noncommutative Yang-Mills theory on $\R^{2,2}$ reads
\begin{equation}\label{eq:ncymact}
  S_{\text{ncYM}} = -{\frac{1}{2g_{\text{YM}}^2}} \int {\mathrm d}^4
    x \; \text{tr}_{u(N)} F_{\mu\nu}\star F^{\mu\nu}.
\end{equation}
Here, $F_{\mu\nu}=\pa_\mu A_\nu - \pa_\nu A_\mu + [A_\mu
\stackrel{\star}{,} A_\nu]$. The self-duality equations in $x^\mu$-%
coordinates are given by
\begin{equation}
  F_{12}=F_{34}, \quad F_{13}=F_{24}\quad\text{and}\quad F_{14}=-F_{23}.
    \label{eq:sdEukl}
\end{equation}
Due to the Bianchi identities for $F_{\mu\nu}$, each solution to~%
(\ref{eq:sdEukl}) will also be a solution to the equations of motion of
noncommutative Yang-Mills theory.

\subsection{Moyal-Weyl map and operator formalism}\label{sec:operator}
\noindent
The Moyal-Weyl map provides us with the possibility to switch between two
equivalent noncommutative formalisms. The noncommutativity of the
configuration space may be captured by deforming the multiplication law
for functions (the Moyal-Weyl- or $\star$-product formalism), which in
turn are defined over a commutative space. Equivalently, we may pass to
the operator formalism, which often simplifies calculations considerably.

\noindent
{\bf Fock space.} In the operator formalism, the coordinates $x^{\mu}$
become operator-valued, thus satisfying $[\hat{x}^{\mu},\hat{x}^{\nu}]=
\ic\th^{\mu\nu}$. More specifically, the commutation relations among
the coordinates $(x,y,\tilde{t},-t)$ are:
\begin{subequations}
\begin{eqnarray}
  \big[\hat{x}, \hat{t} \big] \; = \; \big[ \hat{y},\hat{t} \big] \!
    & = & \! \big[ \hat{x}, \hat{\tilde{t}}\, \big] \; = \; \big[\hat{y},
    \hat{\tilde{t}}\,\big] = 0, \\
  \big[\hat{x}, \hat{y}\big]  \; = \; \ic\th & \Rightarrow & \left[
    \hat{z}^1, \hat{\zb}^1 \right] = 2 \th , \\
  \big[\hat{t},\hat{\tilde{t}}\big] \; = \; \ic \tht &
    \Rightarrow & \left[ \hat{z}^2, \hat{\zb}^2 \right] = 2\tht.
\end{eqnarray}
\end{subequations}
The last two lines lead us to construct creation and annihilation
operators (for $\th,\tht>0$):
\begin{subequations} \label{eq:aop}
\begin{eqnarray}
  a_1 & := & \frac{1}{\sqrt{2\th}} \, \hat{z}^1, \quad a_2 :=
    \frac{1}{\sqrt{2\tht}} \, \hat{z}^2,
    \label{eq:annop}\\
  a_1^{\dag} & := & \frac{1}{\sqrt{2\th}} \, \hat{\zb}^1, \quad
    a_2^{\dag} := \frac{1}{\sqrt{2\tht}} \, \hat{\zb}^2.
\end{eqnarray}
\end{subequations}
These operators act, as usual, in a Fock space ${\cal H}$ constructed
from the action of the two creation operators $a_1^{\dag}, a_2^{\dag}$
on the vacuum $|0,0 \>$. We introduce an orthonormal basis for
${\cal H}$, i.e., $\{|n_1, n_2\> ; n_1,n_2 \in {\mathbb N}_0 \}$
subject to
\begin{eqnarray*}
  & & N_i|n_1, n_2 \> := a_i^{\dag}a_i |n_1, n_2 \> = n_i |n_1, n_2\> ,
    \quad i \in \{1,2\}, \\
  a_1 |n_1, n_2 \> &=& \sqrt{n_1} |n_1 -1, n_2 \> , \quad a_1^{\dag}
    |n_1, n_2 \> = \sqrt{n_1 +1} |n_1 +1, n_2 \> , \\
  a_2 |n_1, n_2 \> &=& \sqrt{n_2} |n_1, n_2 -1 \> , \quad a_2^{\dag}
    |n_1, n_2 \> = \sqrt{n_2 +1} |n_1, n_2 +1 \> .
\end{eqnarray*}
In the case $\tht=0$ we can only introduce $a_1$ and $a_1^\dag$; ${\cal
H}$ will then be a one-oscillator Fock space.

\noindent
{\bf Moyal-Weyl map.} It can be shown that there exists a bijective
map, which maps functions $f(z^i,\zb^i)$ (also called Weyl symbols)
to operators $\hat{f}:=O_f(a_i,a_i^{\dag})$ (cf.\ e.g.\
\cite{Harvey:2001yn,Gross:2000ss}):
\begin{eqnarray}
  f(z^i,\zb^i) \mapsto O_f(a_i,a_i^{\dag}) & = &
    - \int \frac{{\mathrm{d}}^2 k_1 {\mathrm{d}}^2 k_2}{(2 \pi)^4}
    {\mathrm{d}}^2 z^1 {\mathrm{d}}^2 z^2 \\ \nonumber
  & \times & f(z^i,\zb^i) {\mathrm{e}}^{- \ic\{ \bar{k}_1(\sqrt{2
    \th}a_1-z^1) + k_1 (\sqrt{2 \th}a_1^{\dag}-\zb^1) +
    \bar{k}_2(\sqrt{2\tht}a_2-z^2) + k_2 (\sqrt{2\tht}a_2^{\dag}
    -\zb^2)\}},
\end{eqnarray}
where $\int \frac{{\mathrm{d}}^2 k_1 {\mathrm{d}}^2 k_2}{(2 \pi)^4}
{\mathrm{d}}^2 z^1 {\mathrm{d}}^2 z^2 :=\int \frac{{\mathrm{d}}k_1
{\mathrm{d}}\bar{k}_1}{(2 \pi)^2} {\mathrm{d}}z^1 {\mathrm{d}}\zb^1
\int \frac{{\mathrm{d}}k_2 {\mathrm{d}}\bar{k}_2}{(2 \pi)^2}
{\mathrm{d}}z^2 {\mathrm{d}}\zb^2$. Note that this formula implies an
ordering prescription, the so-called Weyl ordering. The inverse
transformation is given by:
\begin{eqnarray}
  O_f(a_i,a_i^{\dag}) & \mapsto & f(z^i,\zb^i) =
    4 \pi^2 \th \tht \int \frac{{\mathrm{d}}^2 k_1 {\mathrm{d}}^2
    k_2}{(2 \pi)^4}  \\ \nonumber
  & \times & {\mathrm{Tr}}_{\cal H} \left[ O_f(a_i,a_i^{\dag})
    {\mathrm{e}}^{\ic\{ \bar{k}_1(\sqrt{2\th}a_1-z^1)+ k_1
    (\sqrt{2 \th}a_1^{\dag}-\zb^1)+\bar{k}_2(\sqrt{2\tht}a_2 -
    z^2) + k_2 (\sqrt{2\tht}a_2^\dag -\zb^2)\}} \right].
    \label{eq:invMW}
\end{eqnarray}
It is understood that, under the Moyal-Weyl map,
\begin{equation}
  f \star g \mapsto \hat{f}\hat{g}.
\end{equation}
Also, an integral $\int {\mathrm{d}}^4 x$ over the configuration
space becomes a trace ${\mathrm{Tr}}_{\cal H}$ over the Fock space
${\cal H}$ (modulo pre-factors) and derivatives are mapped to
commutators, e.g.,
\begin{equation}\label{eq:trans1}
  \pa_x f  \mapsto \frac{\ic}{\th} \big[\hat{y}, \hat{f}\big] ,
  \quad \pa_{z^1} f \mapsto 
  -\frac{1}{\sqrt{2\th}} \big[ a_1^\dag, \hat{f} \big],
\end{equation}
and analogously for the other possible combinations.
From now on, we will work in the operator formalism; exceptions
will be mentioned explicitly. In order to slenderize the notation,
hats will be omitted everywhere.

\section{Dressing approach}\label{sec:dress}
\noindent
As explained in appendix~\ref{sec:twistor}, exact solutions to the
self-duality equations~(\ref{eq:sdEukl}) can be constructed by means
of an associated linear system. Solutions to this linear system will be
obtained via the so-called {\sl dressing method}. It was originally
invented for commutative integrable models as a solution generating
technique to construct solutions to the equations of motion (see,
e.g.~\cite{Zakharov:pp, Forgacs:1983gr, Ward:ie}). New solutions
can be constructed from a simple vacuum seed solution by recursively
applying a dressing transformation. It was shown in~%
\cite{Lechtenfeld:2001aw} that the dressing approach can easily be
extended to noncommutative models. In the following we will apply
such an extension of the dressing method to construct solutions for
the Lax pairs related to the self-duality equations of ncYM on
$\R^{2,2}$.

\subsection{Unitary gauge}\label{sec:unitary}
\noindent
{\bf Lax pair.} Let us start the discussion by considering the Lax
pair given in terms of real isotropic coordinates~\cite{Ivanova:rc}:
\begin{subequations}\label{eq:lax1}
\begin{eqnarray}
  (\z\pa_\vt + \pa_u) \psi & = & -(\z A_\vt + A_u)
    \psi, \label{eq:lax1a} \\
  (\z\pa_v - \pa_\ut) \psi & = & -(\z A_v - A_\ut)
    \psi,  \label{eq:lax1b}
\end{eqnarray}
\end{subequations}
where $A=(A_\mu)$ is the antihermitean gauge potential for the self-%
dual field strength $F=(F_{\mu\nu})$ and $\ps\in GL(N,\C)$.\footnote{For
a detailed discussion concerning the appearance of the complexified gauge
group, we refer to \cite{Lerner:ag, Popov:1998pc}.} As shown in appendix~%
\ref{sec:twistor}, $\psi$ may be chosen to satisfy the following reality
condition:
\begin{equation}\label{eq:reality1}
  \psi(u,v,\ut,\vt,\z)[\psi(u,v,\ut,\vt,\bar{\z})]^{\dag}=1.
\end{equation}
The compatibility conditions for the linear equations~(\ref{eq:lax1})
are given by the self-duality equations expressed in real isotropic
coordinates:
\begin{subequations}\label{eq:realsd}
\begin{eqnarray}
  F_{u\ut} & = & 0, \label{eq:realsd1} \\
  F_{uv} + F_{\ut\vt} & = & 0, \label{eq:realsd2} \\
  F_{v\vt} & = & 0. \label{eq:realsd3}
\end{eqnarray}
\end{subequations}

If we require
\begin{equation}\label{eq:asympt}
  \psi(u,v,\ut,\vt,\zeta \rightarrow 0) =
     g_1^{-1}(u,v,\ut,\vt) + O(\zeta)
\end{equation}
for some $U(N)$ matrix $g_1$ and
\begin{subequations}\label{eq:asympt2}
\begin{eqnarray}
  A_u & = & g_1^{-1} \pa_u g_1^{\phantom{\dag}}, \\
  A_\ut & = & g_1^{-1} \pa_\ut g_1^{\phantom{\dag}},
\end{eqnarray}
\end{subequations}
then eqs.~(\ref{eq:lax1}) in the limit $\z\to 0$ are identically
satisfied~\cite{Yang:1977zf}. Thus, solving~(\ref{eq:lax1}) (without
knowing the gauge fields explicitly, simply by exploiting the
asymptotics of $\psi$) amounts to solving~(\ref{eq:realsd1}). In the
limit $\z\to\infty$, we can read off from~(\ref{eq:lax1}) that
\begin{subequations} \label{eq:Av}
\begin{eqnarray}
  A_\vt & = & g_2^{-1} \pa_\vt g_2^{\phantom{\dag}}, \\
  A_v & = & g_2^{-1} \pa_v g_2^{\phantom{\dag}},
\end{eqnarray}
\end{subequations}
where $g_2^{-1}:= \psi(u,v,\ut,\vt,\z=\infty)\in U(N)$; clearly,~%
(\ref{eq:Av}) solves eq.~(\ref{eq:realsd3}).

\noindent
{\bf Gauge fixing.} Note that we can choose a gauge in which $A_v$
and $A_\vt$ vanish: Consider the gauge transformation
\begin{equation}\label{eq:gtunitary}
  \psi\mapsto\psi' := g_2 \psi .
\end{equation}
Its action on the gauge field yields
\begin{subequations}
\begin{eqnarray}
  A_\vt\mapsto A_\vt' \! & = & \! g_2^{\phantom{\dag}}
    A_\vt g_2^{-1} + g_2^{\phantom{\dag}} \pa_\vt
    g_2^{-1} = 0, \\
  A_v\mapsto A_v' & = & g_2^{\phantom{\dag}} A_v g_2^{-1} +
    g_2^{\phantom{\dag}}\pa_v g_2^{-1} = 0;
\end{eqnarray}
\end{subequations}
this is equivalent to $\ps'(u,v,\ut,\vt,\z=\infty)=1$. For
the remaining components we find
\begin{subequations}\label{eq:gfphi}
\begin{eqnarray}
  A_\ut' & = & \U^{-1} \partial_\ut \U, \\
  A_u' & = & \U^{-1} \partial_u \U,
\end{eqnarray}
\end{subequations}
with $\U^{-1}:=g_2 g_1^{-1}=\psi'(u,v,\ut,\vt,\z=0)$
(Yang prepotential, cf.~\cite{Ivanova:2000zt}). This gauge is called
(unitary) Yang gauge.

The gauge-fixed linear equations read
\begin{subequations}\label{eq:gflax1}
\begin{gather}
  (\z\pa_\vt + \pa_u ) \psi' = -A_u' \psi',
    \label{eq:gflax1a} \\
  (\z\pa_v - \pa_\ut) \psi' = A_\ut' \psi'.
    \label{eq:gflax1b}
\end{gather}
\end{subequations}
Moreover, since $g_2 \in U(N)$, the reality condition~(\ref{eq:reality1})
is ``preserved'' under~(\ref{eq:gtunitary}):
\begin{equation}\label{eq:gfreality1}
  \psi'(u,v,\ut,\vt,\z)[\psi'(u,v,\ut,
    \vt,{\bar{\z}})]^{\dag} = g_2^{\phantom{\dag}} g_2^{\dag}
    = 1.
\end{equation}
In the following we shall omit the primes on the gauge transformed
quantities. Using the above expressions~(\ref{eq:gfphi}) for
$A'_\ut$ and $A'_u$, the remaining self-duality equation~%
(\ref{eq:realsd2}) in this gauge takes the form
\begin{equation}
  \pa_v(\U^{-1}\pa_u \U) + \pa_\vt
    (\U^{-1} \pa_\ut\U) = 0 . \label{eq:gfsdeq}
\end{equation}

\noindent
{\bf Action functional.} Let us introduce an antisymmetric rank two
tensor $\o^{\mu\nu}$ with components $\o^{yt}=-\o^{ty}=-1$,
$\o^{x\tilde{t}}=-\o^{\tilde{t}x}=-1$. Then $\o_{\mu\nu}$ coincides with
$\bar{f}^2_{\mu\nu}$, the analogue to the 't~Hooft tensor in 2+2
dimensions introduced in~\cite{Ivanova:rc}; it is anti-self-dual.
One can interpret $\o=\frac{1}{2}\o_{\mu\nu} dx^\mu\wedge dx^\nu$ as
the K\"ahler form w.r.t.\ the complex structure $\tilde{J}=-\left(
\begin{smallmatrix} 0 & \s^3 \\ \s^3 & 0\end{smallmatrix}\right)$
on $\R^{2,2}$ ($\s^3$ denotes the third Pauli matrix). In $x^\mu$-%
coordinates, we can rewrite eq.~(\ref{eq:gfsdeq}) as
\begin{equation}
  (\eta^{\mu\nu}-\o^{\mu\nu}) \pa_\mu(\U^{-1}\pa_\nu \U)=0.
    \label{eq:gfsdeqwo}
\end{equation}
In contrast to the metric $\eta_{\mu\nu}$, the K\"ahler form is not
invariant under $SO(2,2)$ rotations; it therefore breaks the rotational
invariance of the equation of motion even in the commutative case. A
straightforward computation shows that this is the equation of motion
for the Nair-Schiff type action~\cite{Nair:1991ab, Losev:1995cr}
\begin{equation}
  S = -\frac{1}{2}\int_{\R^{2,2}} d^4 x \,\eta^{\mu\nu}\tr
    (\pa_\mu \U^{-1} \pa_\nu\U) - \frac{1}{3}\int_{\R^{2,2}\times [0,1]}
    \o\wedge\tr(\tilde{A}\wedge \tilde{A}\wedge \tilde{A}).
    \label{eq:NSact}
\end{equation}
Here the gauge potential $A=\U^{-1} d\U$ and the K\"ahler form $\o$ have
nonvanishing components only along $\R^{2,2}$; in the Wess-Zumino term,
$\tilde{A}=\tilde{\U}^{-1} d\tilde{\U}$ is defined via a homotopy
$\tilde{\U}$ from a fixed element $\U_1$ from the homotopy class of $\U$
to $\U$, i.e., $\tilde{\U}(0)=\U_1$, $\tilde{\U}(1)=\U$. Star products are
implicit. Note that the variation w.r.t.\ $\tilde{\U}$ of the Wess-Zumino
term is a total divergence. An ``energy-momentum'' tensor can be easily
obtained from this action; however, we do not want to embark on a
discussion whether it can serve to give a sensible definition of energy
or momentum in 2+2 dimensions. As a simplification, we will sometimes
nevertheless speak of soliton solutions if we can verify that the
solutions have finite energy in 2+1 dimensional subspaces at asymptotic
times.

\noindent
{\bf Dressing approach and ansatz.} Note that, due to~%
(\ref{eq:gfreality1}), eq.~(\ref{eq:gflax1}) can be rewritten as
\begin{subequations}\label{eq:laxrw}
\begin{gather}
  \psi (\z\pa_\vt + \pa_u ) \psi^\dag = A_u,
    \label{eq:laxrw1} \\
  \psi (\z\pa_v - \pa_\ut) \psi^\dag = -A_\ut .
    \label{eq:laxrw2}
\end{gather}
\end{subequations}
It is possible to solve the gauge-fixed linear equations~%
(\ref{eq:laxrw}) without knowing $A_\ut$ and $A_u$ explicitly,
simply by fixing the pole structure\footnote{A nontrivial $\psi(\z)$
cannot be holomorphic in $\z$, since $\z\in\C P^1$, which is compact.}
of $\psi$ in such a way that the left hand sides of (\ref{eq:laxrw}) are
independent of~$\z$. Inserting an ansatz for~$\psi$, we obtain conditions
on its residues which can be solved. From the solution $\psi$, we may
determine $A_\ut$ and $A_{u}$ via eqs.~(\ref{eq:asympt}) and~%
(\ref{eq:asympt2}). Suppose we have constructed a seed solution
$\psi_0$ by solving some appropriate (gauge-fixed) linear equations,
in the present case eqs.~(\ref{eq:gflax1}). Then we can look for a new
solution of the form
\begin{equation}
  \psi_1 = \chi_1 \psi_0 \quad \mathrm{with} \quad
  \chi_1 = 1 + \frac{\mu_1-\mb_1}{\z-\mu_1} P_1,
\end{equation}
where $\mu_1\in{\mathbb{H}}_{\,-}$ (lower half plane) is a complex
constant and where $P_1(u,v,\ut,\vt)$ is an $N \times N$
matrix independent of $\z$. It can be shown that $\mu_1$ may be
interpreted as a modulus parametrizing the velocity of the lump solution
(see e.g.~\cite{Ward:ie, Lechtenfeld:2001aw} for the 2+1 dimensional
case).

Let us start from the vacuum seed solution $\psi_0 = 1$ (the
corresponding gauge potential vanishes). The reality condition~%
(\ref{eq:gfreality1}) for $\psi_1$ is satisfied if we choose $P_1$ to
be a hermitean projector, i.e., $(P_1)^2 = P_1$ and $(P_1)^{\dag} =
P_1$.\footnote{This is the simplest solution to the algebraic conditions
on $P_1$ emerging from the reality condition~(\ref{eq:gfreality1}).}
The transformation $\psi_0 \mapsto \psi_1$ is called {\sl dressing}.
An $m$-fold repetition of this procedure yields
\begin{equation}\label{eq:multansatz}
  \psi_m = \prod_{p=1}^m \left(1+\frac{\mu_p-\mb_p}{\z-\mu_p}
    P_p \right),
\end{equation}
corresponding to an $m$-soliton type configuration if all $\mu_p\in
{\mathbb{H}}_{\,-}$. For~(\ref{eq:multansatz}), the reality condition~%
(\ref{eq:gfreality1}) is automatically satisfied if we choose the $P_p$
to be hermitean (not necessarily orthogonal) projectors. We will see
below that taking all $\mu_p$ to be mutually different will lead us to
interacting plane wave and non-interacting solitons, whereas second-order
poles in~(\ref{eq:multansatz}) (i.e., $\mu_i=\mu_j$ for some $i \neq j$)
entail scattering in soliton-like configurations.

\noindent
{\bf First-order pole ansatz.}
For now, let us restrict to an ansatz~(\ref{eq:multansatz}) containing
only first-order poles in $\z$, i.e., choose all $\mu_p$ to be mutually
different. Then, performing a decomposition into partial fractions, we
can rewrite the multiplicative ansatz~(\ref{eq:multansatz}) in the
additive form
\begin{equation}\label{eq:addansatz}
  \psi_m = 1 + \sum_{p=1}^m \frac{R_p}{\z-\mu_p}.
\end{equation}
The $N \times N$ matrices $R_p(u,v,\ut,\vt)$ are
constructed from multiplicative combinations of the $P_p$; as in~%
\cite{Lechtenfeld:2001aw}, we take the $R_k$ to be of the form
\begin{equation}
  R_p = \sum_{l=1}^m T_l \Gamma^{lp} T_p^{\dag}, \label{eq:RkTk}
\end{equation}
where the $T_l(u,v,\ut,\vt)$ are $N \times r$ matrices
and the $\Gamma^{lp}(u,v,\ut,\vt)$  are $r \times r$
matrices for some $r \geq 1$. The ansatz~(\ref{eq:addansatz}) has to
satisfy the reality condition~(\ref{eq:gfreality1}). Since the right hand
side of the latter is independent of $\z$, the poles on the left hand
side must be removable. Therefore we should equate the corresponding
residues at $\z=\mb_k$ and $\z=\mu_k$ of the left hand side to zero.%
\footnote{In fact, the equation for $\z=\mb_k$ is the hermitean adjoint
to the equation for $\z=\mu_k$. In general, this will hold for any two
residue equations if the points are related by complex conjugation (or,
for $\l$ from section~\ref{sec:hermitean}, by the mapping~%
(\ref{eq:sdef})).} This yields
\begin{equation}
  \left(1-\sum_{p=1}^m \frac{R_p}{\mu_p-\mb_k}\right)T_k = 0.
\end{equation}
These algebraic conditions on $T_k$ imply that the $\Gamma^{lp}$ invert
the matrices
\begin{equation}\label{eq:inverse1}
  \widetilde{\Gamma}_{pk}:= {\frac{T_p^{\dag}T_k}{\mu_p - \mb_k}}, \quad
    \text{i.e., }\sum_{p=1}^m \Gamma^{lp}\widetilde{\Gamma}_{pk}
    = \de_k^l.
\end{equation}

Furthermore, our ansatz should satisfy the gauge-fixed linear equations~%
(\ref{eq:laxrw}). Putting to zero the residues of the left hand sides of~%
(\ref{eq:laxrw}) at $\z=\mu_k$ and $\z=\mb_k$, we learn that
\begin{subequations}\label{eq:rkcond}
\begin{eqnarray}
  \left(1-\sum_{p=1}^m \frac{R_p}{\mu_p - \mb_k} \right)(\mb_k \pa_\vt+
    \pa_u) R_k^{\dag} = 0,\\
  \left(1-\sum_{p=1}^m \frac{R_p}{\mu_p - \mb_k} \right)(\mb_k \pa_v -
    \pa_\ut) R_k^{\dag} = 0,
\end{eqnarray}
\end{subequations}
Thus, we may define new isotropic coordinates (note that $\mu_k$ is
complex) $w_k^1$ and $w_k^2$ in the kernel of the differential operators
in~(\ref{eq:rkcond}):
\begin{subequations} \label{eq:comov1}
\begin{eqnarray}
  w_k^1 := \mb_k^{-1} \vt - u \quad & \text{and} & \quad
    w_k^2 := \mb_k^{-1} v +\ut \\
  \Rightarrow \quad \wb_k^1= \mu_k^{-1} \vt - u
    \quad & \text{and} & \quad \wb_k^2= \mu_k^{-1} v +\ut.
\end{eqnarray}
\end{subequations}
The Lax operators can be written as antiholomorphic vector fields in
terms of these new isotropic coordinates\footnote{In general,
$\bar{L}^1(\z):=\z\pa_\vt+\pa_u$ and $\bar{L}^2(\z):=\z\pa_v-
\pa_\ut$ correspond to the antiholomorphic vector fields introduced
in appendix~\ref{sec:twistor} (in the coordinates $u,v,\ut,\vt$).
Defining coordinate functions $w^1(\z):=\z^{-1}\vt-u$, $w^2(\z):=
\z^{-1}v+\ut$ in their kernel, we may write $\bar{L}_k^{1,2}=
\bar{L}^{1,2}(\z=\mb_k)$ and $w_k^{1,2}=w^{1,2}(\z=\mb_k)$. Furthermore
we have $\bar{L}^{1,2} (\z)=\bar{\z}^{-1}(\z-\bar{\z})\pa_{\wb^{1,2}
(\bar{\z})}$.}
\begin{subequations}\label{eq:antilax}
\begin{align}
  \bar{L}_k^1 := \mb_k \pa_\vt+ \pa_u & = \mu_k^{-1}(\mb_k - \mu_k)
    \frac{\pa}{\pa\wb_k^1}, \\
  \bar{L}_k^2 := \mb_k \pa_v -\pa_\ut & = \mu_k^{-1}(\mb_k - \mu_k)
    \frac{\pa}{\pa\wb_k^2}.
\end{align}
\end{subequations}
As long as $T_k$ is in the kernel of $\bar{L}_k^1$ and $\bar{L}_k^2$,
all functions $R_k$ from~(\ref{eq:RkTk}) automatically solve eqs.~%
(\ref{eq:rkcond}). Thus, special solutions to~(\ref{eq:rkcond}) are
given by~(\ref{eq:RkTk}) with arbitrary differentiable functions
$T_k(w_k^1,w_k^2)$, i.e., $\pa_{\wb^1_k} T_k = 0 = \pa_{\wb^2_k} T_k$
(for each $k=1,\ldots,m$). By inserting such $T_k$ into
\begin{equation}
  \U^{-1} = \psi(u,v,\ut,\vt,\z=0) = 1 - \sum_{l,p=1}^m
    \frac{T_l \Gamma^{lp} T_p^\dag}{\mu_p}, \label{eq:ominv}
\end{equation}
explicit expressions for $A_u, A_\ut$ can be derived from~%
(\ref{eq:gfphi}) and~(\ref{eq:inverse1}).

\subsection{Dimensional reduction to 2+1 dimensions}\label{sec:dimredux}
\noindent
{\bf Dimensional reduction.}
In order to establish the connection between the solutions obtained
above and previous work carried out in 2+1 dimensions\footnote{To
recover the linear system of \cite{Lechtenfeld:2001aw}, we need to
choose the unitary gauge for the linear equations, i.e., Lax pair~%
(\ref{eq:lax1}) as discussed in section~\ref{sec:unitary}.} (see~%
\cite{Lechtenfeld:2001aw}--\cite{Wolf:2002jw}) we have to perform a
dimensional reduction. This can be done by imposing the condition
that all fields are independent of one of the time coordinates in
$\R^{2,2}$. As a consequence, we may put $\tilde{\th}=0$. To be
precise, let us impose
\begin{equation}\label{eq:tindcond}
  \pa_{\tilde{t}} T_k = 0.
\end{equation}
We switch to the complex isotropic coordinates introduced in~%
(\ref{eq:comov1}). Using~(\ref{eq:realder}), we can reexpress
$\frac{\pa}{\pa\tilde{t}}$ as
\begin{equation}
  \frac{\pa}{\pa\tilde{t}} = \frac{1}{2} \left\{ \mb_k^{-1} \frac{\pa}
    {\pa w_k^1} + \mu_k^{-1} \frac{\pa}{\pa\wb_k^{1}} + \frac{\pa}
    {\pa w_k^2} + \frac{\pa}{\pa\wb_k^2} \right\}.
\end{equation}
As derived in section~\ref{sec:unitary}, eqs.~(\ref{eq:rkcond}) are solved
by matrices $T_k$ independent of $\wb_k^1$ and $\wb_k^2$; therefore~%
(\ref{eq:tindcond}) reads
\begin{equation}
  \left[ \mb_k^{-1} \frac{\pa}{\pa w_k^1} + \frac{\pa}{\pa w_k^2}
    \right] T_k(w_k^1, w_k^2) = 0,
\end{equation}
i.e., $T_k$ can only be a function of
\begin{equation}\label{eq:w_k}
  w_k:= \nu_k (w_k^2 - \mb_k w_k^1) = \nu_k \left( x + \frac{1}{2}
    (\mb_k - \mb_k^{-1})y + \frac{1}{2}(\mb_k + \mb_k^{-1})t \right) ,
\end{equation}
if it is independent of the second time direction. The normalization
constant
\begin{equation}\label{eq:nu}
  \nu_k := \left[ \frac{4 \ic}{\mu_k - \mb_k - \mu_k^{-1} + \mb_k^{-1}}
    \cdot \frac{\mu_k - \mu_k^{-1} -2\ic}{\mb_k-\mb_k^{-1} +2\ic}
    \right]^{1/2}
\end{equation}
has been introduced for later convenience. Note that the ``co-moving''
coordinates $w_k$ become static (i.e., independent of $t$) when choosing
$\mu_k=-\ic$; they ``degenerate'' to the complex coordinates $z^1$ from~%
(\ref{eq:cplx}). Conversely, they can be obtained from the ``static''
coordinates $z^1,\bar{z}^1$ by an inhomogeneous $SU(1,1)$ transformation~%
\cite{Lechtenfeld:2001aw}:
\begin{equation}\label{eq:w}
  \begin{pmatrix} w_k \\ \wb_k \end{pmatrix} = \begin{pmatrix} \cosh\,
    \tau_k & -\text{e}^{\ic\vartheta_k} \sinh\, \tau_k \\ -\text{e}^{-\ic
    \vartheta_k} \sinh\,\tau_k & \cosh \,\tau_k \end{pmatrix}
  \begin{pmatrix} z^1 \\ \zb^1 \end{pmatrix} - \sqrt{2 \th}
  \begin{pmatrix} \b_k \\ \bar{\b}_k \end{pmatrix} t,
\end{equation}
where
\begin{subequations}
\begin{align}
  \b_k & = -\frac{1}{2} (2\th)^{-1/2} \nu_k (\mb_k+\mb_k^{-1}), \\
\intertext{and}
  \cosh\,\tau_k & -\text{e}^{\ic\vartheta_k} \sinh\,\tau_k
    = \nu_k, \quad \text{e}^{\ic\vartheta_k} \tanh\,\tau_k
    = \frac{\mb_k -\mb_k^{-1} -2\ic}{\mb_k -\mb_k^{-1} +2\ic}.
\end{align}
\end{subequations}
Recall that a general solution $T_k$ in 2+2 dimensions is an arbitrary
function of $w_k^1, w_k^2$. Hence, dimensional reduction to 2+1 dimensions
can be accomplished for $T_k$ depending only on $w_k$:
\begin{equation}
  \pa_{\wb^1_k} T_k = 0 = \pa_{\wb^2_k} T_k \quad\text{and}\quad
    \pa_{\tilde{t}} T_k = 0 \qquad \Leftrightarrow \qquad T_k = T_k(w_k).
\end{equation}
The Lax operators acting in this 2+1 dimensional subspace are given by:
\begin{subequations}
\begin{eqnarray}
  \bar{L}_k^{1} & = & \mb_k \pa_{\tilde{t}} - \mb_k \pa_x + \pa_u
    \stackrel{2+1}{\longrightarrow} - \bar{\nu}_k (\mb_k - \mu_k)
    \pa_{\wb_k}, \\
  \bar{L}_k^{2} & = & \mb_k \pa_v - \pa_{\tilde{t}} - \pa_x
    \stackrel{2+1}{\longrightarrow} \bar{\nu}_k \mu_k^{-1}(\mb_k - \mu_k)
    \pa_{\wb_k}.
\end{eqnarray}
\end{subequations}
This exactly matches the results of~\cite{Lechtenfeld:2001aw}.

Note that an alternative reduction can be done if $T_k$ is independent
of $t$ (but depends on all other coordinates):
\begin{equation}
  \frac{\pa}{\pa t} = -\frac{1}{2} \left\{ \frac{\pa}{\pa w_k^1} +
    \frac{\pa}{\pa\wb_k^1} - \mb_k^{-1} \frac{\pa}{\pa w_k^2}
    - \mu_k^{-1} \frac{\pa}{\pa\wb_k^2}\right\}.
\end{equation}
Then, the condition $\pa_t T_k = 0$ and eqs.~(\ref{eq:rkcond}) are
satisfied for $T_k = T_k(\wt_k)$ with
\begin{equation}
  \wt_k:= \nu_k (w_k^{1} + \mb_k w_k^2) = \nu_k \left( -y
    +\frac{1}{2} (\mb_k -\mb_k^{-1})x +\frac{1}{2} (\mb_k + \mb_k^{-1})
    \tilde{t} \right). \label{eq:wt_k}
\end{equation}
Note that a ``boost'' transformation analogous to~(\ref{eq:w}) can be
found for the coordinates $\wt_k$:
\begin{equation}\label{eq:wtilde}
  \begin{pmatrix} \wt_k \\ \ov{\wt}_k \end{pmatrix} = \ic \begin{pmatrix}
    \cosh\,\tau_k & \text{e}^{\ic\vartheta_k} \sinh\,\tau_k \\
    -\text{e}^{-\ic\vartheta_k} \sinh\,\tau_k & -\cosh\,\tau_k
    \end{pmatrix} \begin{pmatrix} z^1 \\ \zb^1 \end{pmatrix} -\sqrt{2\th}
  \begin{pmatrix} \b_k \\ \bar{\b}_k \end{pmatrix} \tilde{t}.
\end{equation}
Such $T_k(w_k)$ or $T_k(\wt_k)$ lead to $\U$ which are given by~%
(\ref{eq:ominv}) and do not depend on $\tilde{t}$ or $t$, respectively.

\noindent
{\bf Map to operator formalism.}
If we translate the co-moving coordinates $w_k$ and $\wt_k$ into the
operator formalism, this yields co-moving creation and annihilation
operators:
\begin{subequations}\label{eq:opcdef}
\begin{eqnarray}
  \wh_k^\dag = \wbh_k & \Rightarrow & \big[ \wh_k,\wbh_k\big] = 2\th , \\
  \widehat{\wt}_k^\dag = \widehat{\ov{\wt}}_k & \Rightarrow &
    \big[ \widehat{\wt}_k,\widehat{\ov{\wt}}_k \big] = 2\th .
\end{eqnarray}
\end{subequations}
Note that, in general, the commutators between $\wh_k$ and
$\widehat{\wt}_k$ will not vanish. Therefore, derivatives with respect to
$w_k$ and $\wb_k$ translate into commutators (cf.~(\ref{eq:trans1})) of
the simple form
\begin{equation}
  2\th\pa_{w_k}= - \big[ \wbh_k,\, .\big], \quad 2\th\pa_{\wb_k}
    = \big[ \wh_k,\, .\big], 
\end{equation}
only when acting on functions of $\wh_k$ and $\wbh_k$. Analogous
statements hold for derivatives with respect to $\wt_k, \ov{\wt}_k$.
In this framework, the transformations~(\ref{eq:w}) and~(\ref{eq:wtilde})
may be interpreted as Bogoliubov transformations relating $\hat{z}^1$ and
$\hat{\zb}^1$ to the operators in~(\ref{eq:opcdef})~%
\cite{Lechtenfeld:2001aw}.

\noindent
{\bf Energy.}
In 2+1 dimensions it is possible to define the notion of energy in a
straightforward manner and to show that it is conserved. Dimensional
reduction of the Nair-Schiff type action~(\ref{eq:NSact}) leads to the
action for a modified noncommutative sigma model in 2+1 dimensions as
presented in~\cite{Lechtenfeld:2001aw}. From this, an energy-momentum
tensor can easily be derived:
\begin{equation}
  T_{cd} = (\de_c^a \de_d^b - \frac{1}{2} \eta_{cd}\eta^{ab})\tr(\pa_a
    \U^{-1} \pa_b \U), \label{eq:emtred}
\end{equation}
$a$, $b$, $c$ and $d$ running over $x, y$, and $t$. For the proof that
$T_{cd}$ is divergence-free we need to apply the equation of motion
\begin{equation}
  (\eta^{ab}-\o^{ab}) \pa_a(\U^{-1}\pa_b\U) = 0 \label{eq:sdeomtpo}
\end{equation}
obtained by dimensional reduction (by imposing $\pa_{\tilde{t}}(\U^{-1}
\pa_b\U)=0$) from eq.~(\ref{eq:gfsdeqwo}). Using the explicit form of
$\o^{\mu\nu}$, it is obvious that one can rewrite eq.~(\ref{eq:sdeomtpo})
as
\begin{equation}
  (\eta^{ab}+V_c\ve^{cab}) \pa_a(\U^{-1}\pa_b\U) = 0,
\end{equation}
where $(V_c)=(V_x,V_y,V_t)=(1,0,0)$ manifestly breaks Lorentz-invariance
even in the commutative case. With this, one finds that
$\int d^2 x\,\pa^a T_{at}$ vanishes due to the chosen form of $\o_{ab}$.
For the energy density, one obtains
\begin{equation}
  {\cal E} = T_{tt} = \frac{1}{2} \tr [(\pa_t \U^\dag)\pa_t \U + (\pa_x
    \U^\dag)\pa_x \U + (\pa_y \U^\dag)\pa_y \U] ; \label{eq:enden}
\end{equation}
obviously $\pa_t \int d^2 x\,{\cal E} = 0$.

\noindent
{\bf Nonabelian soliton in 2+1 dimensions.}
As an illustrative example, consider a nonabelian one-soliton ($m=1$) in
2+1 dimensions as described in~\cite{Lechtenfeld:2001aw}. Since
$m=1$, we may start from~(\ref{eq:multansatz}) with $P_1\equiv P=T(T^\dag
T)^{-1}T^\dag$, cf.~(\ref{eq:RkTk}) and~(\ref{eq:inverse1}). For
definiteness, we take the soliton to be embedded into the $xyt$-plane,
i.e., $T_1\equiv T$ is a function of $w_1\equiv w$ (cf.~(\ref{eq:w_k})).
Such a function $T$ trivially solves~(\ref{eq:rkcond}).

Exemplarily, we briefly review a solution corresponding to a moving
$U(2)$ soliton~\cite{Lechtenfeld:2001aw}. Using the inverse Moyal-Weyl
map, we can deduce from the simplest ansatz $T=\left( \begin{smallmatrix}
1 \\ w \end{smallmatrix}\right)$ that
\begin{equation}
  \U_\star = 1 - \frac{\mb-\mu}{\mb} \begin{pmatrix}
    \frac{2\th}{w\wb+\th} & \frac{\sqrt{2\th}w\wb^2}{(w\wb+\th)^2} \\
    \frac{\sqrt{2\th}^{\phantom{\dag}}w^2\wb}{(w\wb+\th)^2} &
    \frac{w\wb+\th}{w\wb+3\th}
  \end{pmatrix} , \label{eq:onesol}
\end{equation}
with the ordinary product between $w$ and $\wb$, solves the self-duality
equation. With the help of~(\ref{eq:enden}), the energy of this
configuration can be shown to be $E=8\pi\cosh\eta\sin\varphi$ where
$e^{\eta-\ic\varphi}=\mu$.

A remark on the interpretation of solitons in terms of D-branes is in
order: We start out from ncSDYM on a space-time filling D-brane.
If a solution $\psi$ is independent of one coordinate, we are
allowed to compactify and subsequently T-dualize this direction. This
alters the Neumann boundary conditions for open strings living on the
space-time filling branes to Dirichlet boundary conditions.
In this case we therefore consider gauge theory on a D2-brane. Although
there exists no Hodge self-duality condition in such a three-dimensional
gauge theory, we will (in a slight abuse of language) still speak of
solitonic solutions (implicitly referring to the four-dimensional gauge
theory before T-dualization).

Since $\U_\star$ from~(\ref{eq:onesol}) and the corresponding energy
density are independent of $\tilde{t}$, a T-dualization in the
$\tilde{t}$-direction leads to a gauge configuration on a pair of
D2-branes. Taking into account that $\U$ depends only on two variables
$w, \wb$ in three dimensions, we conclude that it corresponds to a
D0-brane moving in the world-volume of two D2-branes.

\subsection{Hermitean gauge}\label{sec:hermitean}
\noindent
{\bf Lax pair.}
Instead of using $\z$, the Riemann sphere $\C P^1$ may alternatively be
parametrized by the variable
\begin{equation}\label{eq:trafo}
  \l = \frac{\z - \ic}{\z + \ic}.
\end{equation}
The map $\z\mapsto\l$ carries the lower half plane in $\z$ to the exterior
of the unit disk $\{|\l|>1\}$ in the $\l$-plane. In terms of $\l$ and the
coordinates $z^1, \zb^1, z^2, \zb^2$ on $\R^{2,2}\cong\C^{1,1}$, the Lax
pair~(\ref{eq:lax1}) becomes\footnote{The conventions for $z^1, z^2$ are
such that for $\mu_k'=\infty$ we obtain holomorphic functions $T$ as
solutions of~(\ref{eq:rkcond2}).}
\begin{subequations}\label{eq:laxpair2}
\begin{eqnarray}
  (\pa_{\zb^1} - \l \pa_{z^2}) \psi = -(A_{\zb^1} - \l A_{z^2}) \psi,
    \label{eq:laxpair2a} \\
  (\pa_{\zb^2} - \l \pa_{z^1}) \psi = -(A_{\zb^2} - \l A_{z^1}) \psi,
    \label{eq:laxpair2b}
\end{eqnarray}
\end{subequations}
and its compatibility conditions are the self-duality equations
\begin{subequations}\label{eq:sdcomplex}
\begin{eqnarray}
  F_{z^1 z^2} & = & 0 , \label{eq:sdcomplex1} \\
  F_{z^1 \zb^1} - F_{z^2 \zb^2} & = & 0, \label{eq:sdcomplexl1} \\
  F_{\zb^1 \zb^2} & = & 0. \label{eq:sdcomplex2}
\end{eqnarray}
\end{subequations}
Here, $\psi$ may be chosen to satisfy the reality condition
\begin{equation}\label{eq:reality2}
  \psi(z^1,\zb^1,z^2,\zb^2,\l) [\psi(z^1,\zb^1,z^2,\zb^2,
    \lb^{-1})]^{\dag} = 1.
\end{equation}

Equations~(\ref{eq:sdcomplex1}) and (\ref{eq:sdcomplex2}) imply
that there exist $g,\tilde{g} \in GL(N,\C)$ such that:
\begin{subequations}
\begin{eqnarray}
  A_{z^1} = g^{-1} \pa_{z^1}  g, & \quad & A_{z^2} =
    g^{-1} \pa_{z^2}  g, \\
  A_{\zb^1} = {\tilde{g}}^{-1} \pa_{\zb^1} \tilde{g}, & \quad &
    A_{\zb^2} = {\tilde{g}}^{-1} \pa_{\zb^2} \tilde{g}.
\end{eqnarray}
\end{subequations}
We read off that a possible choice for $g$ and $\tilde{g}$ is given by
\begin{subequations} \label{eq:psias}
\begin{eqnarray}
  g & := & [\psi(z^i, \zb^i, \l\to \infty)]^{-1} , \\
  \tilde{g} & := & [ \psi(z^i, \zb^i, \l\to 0)]^{-1} .
\end{eqnarray}
\end{subequations}
Since we are using antihermitean generators for the gauge group $U(N)$,
the $GL(N,\C)$-valued fields $g, \tilde{g}$ have to be related:
\begin{equation}\label{eq:antih}
  A_{z^i}^{\dag} = - A_{\zb^i}, \; i = 1,2 \quad \Rightarrow \quad
    \tilde{g} = (g^{\dag})^{-1}.
\end{equation}

\noindent
{\bf Gauge fixing.}
As in section~\ref{sec:unitary}, we can perform a gauge transformation
to set two components of the gauge potential to zero. Contrary to the
(unitary) gauge choice there, in the following we set to zero those
components which are {\sl not} multiplied by the respective spectral
parameter in eqs.~(\ref{eq:laxpair2}).\footnote{In this way, we
facilitate a comparison with~\cite{Lechtenfeld:2002cu}.} Explicitly,
\begin{subequations} \label{eq:Yangg}
\begin{gather}
  \psi' = \tilde{g} \psi, \\
  A_{z^1}' = h^{-1} \pa_{z^1} h, \qquad A_{z^2}' = h^{-1}
    \pa_{z^2} h, \label{eq:Yanggb} \\
  A_{\zb^1}' = 0, \qquad A_{\zb^2}' = 0,
\end{gather}
\end{subequations}
where $h:=g \tilde{g}^{-1} = g g^{\dag} \in GL(N,\C)$ is hermitean. This
gauge is ``asymmetric'', i.e., the gauge potential does not obey~%
(\ref{eq:antih}), but instead it satisfies $(A'_{z^i})^\dag=-h A'_{\zb^i}
h^{-1}-h \pa_{\zb^i} h^{-1}$. After solving~(\ref{eq:sdcomplex}) we are
free to gauge back to a ``symmetric'' gauge, where~(\ref{eq:antih}) is
restored. This is ensured by the hermiticity of $h$, which is the
remnant of~(\ref{eq:antih}) in the asymmetric gauge. We will from now on
work in the asymmetric gauge and omit all primes on the gauge-transformed
quantities.

Now, the gauge-fixed linear equations read
\begin{subequations}\label{eq:gflax2}
\begin{eqnarray}
  (\pa_{\zb^1} - \l\pa_{z^2}) \psi = \l A_{z^2} \psi,
    \label{eq:gflax2a} \\
  (\pa_{\zb^2} - \l\pa_{z^1}) \psi = \l A_{z^1} \psi.
    \label{eq:gflax2b}
\end{eqnarray}
\end{subequations}
Due to~(\ref{eq:antih}), the reality condition~(\ref{eq:reality2})
transforms into\footnote{This is the reason why we call this gauge
hermitean. It coincides with the hermitean gauge introduced in~%
\cite{Yang:1977zf}.}
\begin{equation}\label{eq:gfreality2}
  \psi(\l) [\psi(\lb^{-1})]^{\dag} = \tilde{g}g^{-1} = h^{-1}.
\end{equation}
In the asymmetric gauge the remaining self-duality equation~%
(\ref{eq:sdcomplexl1}) reduces to
\begin{equation}\label{eq:newsd}
  \pa_{\zb^1} (h^{-1}\pa_{z^1} h)-\pa_{\zb^2} (h^{-1}\pa_{z^2} h) = 0.
\end{equation}

\noindent
{\bf First-order pole ansatz.}
Since the reality condition~(\ref{eq:gfreality2}) is different from
the one in the unitary gauge, we are forced to employ a modified
ansatz for $\psi(\l)$. The first-order pole ansatz for $\psi$ takes
the form\footnote{The parameters $\mu_p'$ are the images of $\mu_p$
under (\ref{eq:trafo}). However, the ansatz~(\ref{eq:newpsi}) is not
simply the transform of~(\ref{eq:addansatz}).}
\begin{equation}
  \psi_m(\l)=1+\sum_{p=1}^m {\frac{\l{\widetilde{R}}_p}{\l-\mu'_p}},
    \label{eq:newpsi}
\end{equation}
where
\begin{equation}\label{eq:newr}
  {\widetilde{R}}_p:=-\sum_{q=1}^m \mu'_p T_p \Gamma^{pq} T_q^{\dag}.
\end{equation}
The ``inverse'' matrix $\widetilde{\Gamma}=(\widetilde{\Gamma}_{pk})$,
cf.~(\ref{eq:inverse1}), here reads:
\begin{equation}\label{eq:Gtpkh}
  \widetilde{\Gamma}_{pk} = \mu'_p \frac{T_p^{\dag}T_k}{1-\mu'_p\mb'_k}.
\end{equation}

The matrix-valued function $\psi_m$ should satisfy the linear equations~%
(\ref{eq:gflax2}) and is subject to a reality condition, eq.~%
(\ref{eq:gfreality2}). Again, the requirement that the poles at $\l=
\mb_k'^{-1}$  and $\l=\mu_k'$ of~(\ref{eq:gfreality2}) have to be
removable yields\footnote{Note that $h^{-1}$ is independent of $\l$.}
\begin{equation}\label{eq:tkcond2}
  \bigg( 1-\sum_{p=1}^m  \frac{{\widetilde R}_p}{\mu'_p \mb'_k - 1}
    \bigg) T_k = 0,
\end{equation}
which is solved by~(\ref{eq:newr}) with~(\ref{eq:Gtpkh}). Now we
exploit the pole structure of the Lax pair which, using~%
(\ref{eq:gfreality2}), may be rewritten as
\begin{subequations}
\begin{eqnarray}
  \left[ \left( \frac{1}{\l}\pa_{\zb^1} - \pa_{z^2} \right) \psi
    \right] \psi^\dag & = & A_{z^2} h^{-1} \\
  \left[ \left( \frac{1}{\l}\pa_{\zb^2} - \pa_{z^1} \right) \psi
    \right] \psi^\dag & = & A_{z^1} h^{-1} .
\end{eqnarray}
\end{subequations}
As before, the right hand sides do not feature poles in $\l$, therefore
taking the residue at $\l=\mb_k'^{-1}$ and $\l=\mu_k'$ leads to the
conditions
\begin{subequations}\label{eq:rkcond2}
\begin{eqnarray}
  \bigg( 1-\sum_{p=1}^m \frac{\widetilde{R}_p}{\mu'_p \mb'_k - 1}
    \bigg) (\pa_{\zb^1} - \mb_k'^{-1} \pa_{z^2})\widetilde{R}_k = 0, \\
  \bigg( 1-\sum_{p=1}^m \frac{\widetilde{R}_p}{\mu'_p \mb'_k - 1}
    \bigg) (\pa_{\zb^2} - \mb_k'^{-1} \pa_{z^1})\widetilde{R}_k = 0.
\end{eqnarray}
\end{subequations}
If we define
\begin{subequations} \label{eq:wprdef}
\begin{eqnarray}
  \eta^1(\l) := z^1 + \l\zb^2 \quad & \Rightarrow & \quad
    \eb^1(\lb) = \zb^1 + \lb z^2, \\
  \eta^2(\l) := z^2 + \l\zb^1 \quad & \Rightarrow & \quad
    \eb^2(\lb) = \zb^2 + \lb z^1,
\end{eqnarray}
\end{subequations}
and denote $\eta_k^i:=\eta^i(\l=\mb_k'^{-1})$, $\eb_k^i:=\eb^i(\lb=
\mu_k'^{-1})$, the Lax operators can be written as antiholomorphic
vector fields in these coordinates:
\begin{eqnarray}
  \bar{L}_k^1 & = & \pa_{\zb^1} - \mb_k'^{-1} \pa_{z^2} =
    (1-|\mu_k'|^{-2}) \frac{\pa}{\pa\eb_k^1}, \\
  \bar{L}_k^2 & = & \pa_{\zb^2} - \mb_k'^{-1} \pa_{z^1} =
    (1-|\mu_k'|^{-2}) \frac{\pa}{\pa\eb_k^2}.
\end{eqnarray}
Functions $T_k=T_k(\eta_k^1,\eta_k^2)$ are in the kernel of
$\bar{L}_k^1$ and $\bar{L}_k^2$; therefore $\widetilde{R}_k$ constructed
via~(\ref{eq:newr}) from such $T_k$ automatically satisfy eqs.~%
(\ref{eq:rkcond2}). Due to~(\ref{eq:psias}) and~(\ref{eq:gfreality2}),
we can determine $A_{z^1}$ and $A_{z^2}$ in eq.~(\ref{eq:Yangg}) from
\begin{equation}
  h^{-1} = 1 + \sum_{p=1}^m \widetilde{R}_p .
\end{equation}

\subsection{Relation to string field theory} \label{sec:SFT}
\noindent
In this section, we want to clarify the relation to N=2 string field
theory in the presence of a $B$-field. In the next paragraph, we will
show that the zero-mode part of the string field theory equation of motion
contains the field theory self-duality equation. After that, it will
turn out that, in the Seiberg-Witten limit, an analogous discussion of
the dressing approach~\cite{Lechtenfeld:2002cu} leads to Lax operators
acting only on the oscillator (non-zero mode) part of a string field.%
\footnote{Since no special form of the star product was demanded in~%
\cite{Lechtenfeld:2002cu}, the discussion there is equally well valid
in the case of nonvanishing $B$-field.}

\noindent
{\bf Field theory content of string field theory.}
Let us first briefly show that the gauge-fixed self-duality equation~%
(\ref{eq:newsd}) is contained in the equation of motion of nonpolynomial
string field theory for N=2 strings~\cite{Berkovits:1995ab,
Berkovits:1997pq, Lechtenfeld:2000qj} (for finite $\a'$). Its equation
of motion in the conventions of~\cite{Lechtenfeld:2002cu} reads
\begin{equation}
  \Gtp (\text{e}^{-\P} \Gp \text{e}^\P) = 0 , \label{eq:Berk}
\end{equation}
where $\P$ is a hermitean string field. The operators \Gp and \Gtp are
constituents of a small N=4 superconformal algebra acting on string fields
via contour integration, e.g.,
\begin{equation}
  (\Gp \text{e}^\P)(w) = \oint \frac{dw'}{2\pi\ic}\,\Gp(w')
    \text{e}^\P(w) , \label{eq:contour}
\end{equation}
where the integration contour runs around $w$. Note that, in this
context, $w$ and $w'$ are world-sheet coordinates. All string fields in~%
(\ref{eq:Berk}) are multiplied by the Witten star product; this will be
analyzed more deeply in the next subsection. If we denote the complex
N=2 world-sheet bosons by~$Z^i$ and $\Zb^i$ ($i=1,2$) and their
NSR superpartners as $\psi^{+i}$ and $\psi^{-i}$, the superconformal
generators in~(\ref{eq:Berk}) can be realized as~\cite{Berkovits:1994vy,
Junemann:1999hi}
\begin{equation}
  \Gp = \eta_{i\jb}\psi^{+i} \pa\Zb^j \qquad\text{and}\qquad
  \Gtp = -\ve_{ij}\ps^{+i}\pa Z^j . \label{eq:GpGtp}
\end{equation}
Here, $\eta_{i\jb}$ denotes the (pseudo-)K\"ahler space-time metric with
non-vanishing components $\eta_{1\bar{1}}=-\eta_{2\bar{2}}=1$; for the
antisymmetric tensor we choose the convention that $\ve_{12}=-1$. Taking
into account the bosonic operator product expansions
\begin{equation}
  Z^i(w,\wb) \Zb^j(w',\wb') \sim -2\a'\eta^{i\jb}\ln|w-w'|^2 , \quad
  Z^i(w,\wb) Z^j(w',\wb') \sim 0 , \quad
  \Zb^i(w,\wb) \Zb^j(w',\wb') \sim 0,
\end{equation}
we see that due to~(\ref{eq:contour}) \Gp and \Gtp act as derivatives
on string fields containing only world-sheet bosons. Concretely, the
equations of motion~(\ref{eq:Berk}) for such string fields can be
written as
\begin{equation}
  \psi_0^{+1} \psi_0^{+2} \eta^{i\jb}\pa_{\zb^j}\left(
    \text{e}^{-\P} \pa_{z^i} \text{e}^\P \right) + \ldots
    = 0 . \label{eq:eomzm}
\end{equation}
Here, the bosonic zero-modes~$z^i$ and~$\zb^j$ coincide with the
space-time coordinates used in section~\ref{sec:hermitean};
$\psi_0^{+i}$ denote the zero-modes of $\psi^{+i}$. The dots indicate
the oscillator-dependent part of the equation of motion. The zero-mode
part in~(\ref{eq:eomzm}) coincides with the remaining self-duality
equation~(\ref{eq:newsd}) in the Yang gauge~(\ref{eq:Yangg}) if we
identify~$h(z^i,\zb^i)=e^{\P(z^i,\zb^i)}$ (cf.~(\ref{eq:Yanggb})).

In the same way, the linear equation given in~\cite{Lechtenfeld:2002cu}
includes the field theory Lax pair~(\ref{eq:gflax2}). For $A=
\text{e}^{-\P}\Gp\text{e}^\P$, it can be written as
\begin{eqnarray}
  0 & = & \{\Gp + \l\Gtp + \l A\}\Psi \label{eq:sftLax} \\
    & = & \frac{1}{2} \left\{ \psi_0^{+1}\left( \pa_{\zb^2}-\l\pa_{z^1}
            -\l \text{e}^{-\P}\pa_{z^1} \text{e}^\P \right) + \psi_0^{+2}
            \left( \pa_{\zb^1}-\l\pa_{z^2}-\l\text{e}^{-\P}\pa_{z^2}
            \text{e}^\P \right) + \ldots \right\} \Psi . \nonumber
\end{eqnarray}
Because $\psi_0^{+1}$ and $\psi_0^{+2}$ are mutually independent, the
zero-mode part coincides with~(\ref{eq:gflax2}).

\noindent
{\bf Star product in the Seiberg-Witten limit.}
Now we will scrutinize the Seiberg-Witten limit of string field theory
in a $B$-field background and argue that, in this limit, the above
BRST-like operators \Gp and \Gtp act only on the oscillator-part of
a string field $\text{e}^\P$.\footnote{This argument is along the lines
of~\cite{Witten:2000nz, Schnabl:2000cp}.} To this aim, we switch
to real coordinates $x^\mu$ according to~(\ref{eq:cplx}).
In covariant string field theory, strings are glued with Witten's star
product identifying the left half of the first string with the right
half of the second string. This product is noncommutative even without
a $B$-field background, but in order to make contact with the discussion
of ncSDYM in this paper, we switch on a constant $B$-field.\footnote{For
N=2 strings, the $B$-field must be a (pseudo-)K\"{a}hler two-form~%
\cite{Lechtenfeld:2000nm}.} There are several ways to compute Witten's
star product; the one most suitable for our purposes is the use of an
oscillator representation of the three-vertex ${}_{123}\<V_3|$ joining
two string states $|A\>_1$ and $|B\>_2$ according to
\begin{subequations}
\begin{eqnarray}
  {}_3\<C| & = & {}_{123}\< V_3| |A\>_1 |B\>_2 \\
  \hspace{-1cm}\text{with}\quad {}_{123}\< V_3| & = & \left(
    \frac{3\sqrt{3}}{4} \right)^3 \de(p^{(1)} + p^{(2)} + p^{(3)})
    \< 0| \otimes \< 0| \otimes \< 0|
    \exp(E_\text{mat}), \\ %
  E_\text{mat} & = & \sum_{m,n=0}^\infty \frac{1}{2} \a^{(r)\mu}_n
    N^{rs}_{nm} \a^{(s)\nu}_m G_{\mu\nu} - \frac{\ic}{2}\th_{\mu\nu}
    p^{(1)\mu} p^{(2)\nu} 
    +\sum_{m,n=0}^\infty \frac{1}{2} \ps^{(r)\mu}_n V^{rs}_{nm}
    \ps^{(s)\nu}_m G_{\mu\nu} ,\quad \label{eq:thrvertm}
\end{eqnarray}
\end{subequations}
where $N^{rs}_{nm}$ and $V^{\pm rs}_{nm}$ are the Neumann coefficients~%
\cite{Gross:1986ia} for world-sheet bosons and fermions and
$\a^{(r)\mu}_n$ and $\ps^{\pm(r)\mu}_n$ denote the bosonic and fermionic
oscillators of the $r$-th string, respectively. The open string metric
$G_{\mu\nu}$ was introduced in eq.~(\ref{eq:GBdef}), and we write
$\a_0^\mu=\sqrt{2\a'}p^\mu$. A summation over $r, s=1,2,3$ and over
$\mu,\nu=1,\ldots,4$ is implicit. This expression is valid for N=2
strings and is constructed analogous to~\cite{Schnabl:2000cp,
Bonora:2002iq, Bonora:2002rn}. Note that for N=2 nonpolynomial string
field theory no world-sheet ghosts are needed.

We will now consider the properties of this vertex in the Seiberg-Witten
limit $B\to\infty$ keeping fixed all other closed string parameters. For
this purpose, we set $B=t B_0$ and take $t\to\infty$;\footnote{This limit
is not to be confused with the large time limit in section~%
\ref{sec:scattering}.} then, the effective open string parameters scale
as~\cite{Schnabl:2000cp}
\begin{equation}
  G_{\mu\nu} \sim G_{0\mu\nu} t^2 , \qquad
    \th^{\mu\nu} \sim \th_0^{\mu\nu} t^{-1} .
\end{equation}
In checking the operator product expansions for the N=4 superconformal
algebra, the relations
\begin{equation}
  \ve_{ij} \eta^{j\jb}\ve_{\jb\ib} = \eta_{i\ib} \label{eq:epseta}
\end{equation}
are needed. Since $\eta_{i\jb}$ in eq.~(\ref{eq:GpGtp}) has to be
replaced by $G_{i\jb}$ in the case of a nonvanishing $B$-field, the
``(anti)holomorphic part of the volume element'' $\ve_{ij}$ is changed
to $\epsilon_{ij}$ with the same scaling behavior as $G_{ij}$ (cf.~%
(\ref{eq:epseta})).

For the commutation relations
\begin{subequations}
\begin{eqnarray}
  [\a^\mu_m, \a^\nu_n] & = & m \de_{m+n,0} G^{\mu\nu} , \\
  \big[ x^\mu, x^\nu \big] & = & \ic \th^{\mu\nu} , \\
  \big[ p^\mu, x^\nu \big] & = & -\ic G^{\mu\nu} , \\
  \{ \ps_m^\mu, \ps_n^\nu \} & = & \de_{m+n,0} G^{\mu\nu}
\end{eqnarray}
\end{subequations}
to be invariant in the large $B$-field limit, we have to introduce
rescaled oscillators
\begin{subequations}
\begin{eqnarray}
  \tilde{\a}^\mu_m & = & t \a^\mu_m \quad\text{for}\; m\neq 0,\\
  \tilde{p}^\mu & = & t^{3/2} p^\mu , \\
  \tilde{x}^\mu & = & t^{1/2} x^\mu , \\
  \tilde{\ps}^\mu_m & = & t \ps^\mu_m .
\end{eqnarray}
\end{subequations}
In terms of these modes, the matter part of the three-vertex~%
(\ref{eq:thrvertm}) takes the form
\begin{eqnarray}
  E_\text{mat} & = &\sum_{m,n=1}^\infty \frac{1}{2} \tilde{\a}^{(r)\mu}_n
    N^{rs}_{nm} \tilde{\a}^{(s)\nu}_m G_{0\mu\nu} - \frac{\ic}{2}
    \th_{0\mu\nu}\tilde{p}^{(1)\mu} \tilde{p}^{(2)\nu}
    + \frac{1}{\sqrt{t}} \sum_{n=1}^\infty \sqrt{\frac{\a'}{2}}
    \tilde{\a}^{(r)\mu}_n (N^{rs}_{n0}+N^{rs}_{0n}) \tilde{p}^{(s)\nu}
    G_{0\mu\nu} \nonumber \\
  & & \!\!\!\! {}+\frac{1}{t}\a' \tilde{p}^{(r)\mu} N^{rs}_{00}
    \tilde{p}^{(s)\nu}G_{0\mu\nu}+\sum_{m,n=0}^\infty \frac{1}{2}
    \tilde{\ps}^{(r)\mu}_n V^{rs}_{nm} \tilde{\ps}^{(s)\nu}_m
    G_{0\mu\nu} .
\end{eqnarray}
Note that, for $t\to\infty$, the terms coupling $\a$-oscillators and
momenta~$p$ vanish. Thus, the string star algebra $\cal A$ factorizes
into a zero-momentum part ${\cal A}_0$ spanned by~$\tilde{p}$-,
$\tilde{\a}$-, and $\tilde{\ps}$-oscillators and a space-time part
${\cal A}_1$ generated by $\tilde{x}^\mu$~\cite{Witten:2000nz}. The
star product in ${\cal A}_1$ ``degenerates'' to the usual Moyal-Weyl
product with constant noncommutativity parameter $\th_0$.

To read off the scaling behavior of the BRST-like operators \Gp and
\Gtp, we switch back to complex coordinates (labeled by roman space-time
indices) and exemplarily pick two typical terms from \Gp:
\begin{equation}
  \ps_0^{+i} p^\jb G_{i\jb} = \frac{1}{\sqrt{t}}\tilde{\ps}_0^{+i}
    \tilde{p}^\jb G_{0i\jb} \label{eq:Gp1}
\end{equation}
and
\begin{equation}
  \ps_1^{+i} \a_{-1}^\jb G_{i\jb} = \tilde{\ps}_1^{+i}
    \tilde{\a}_{-1}^\jb G_{0i\jb} . \label{eq:Gp2}
\end{equation}
Eq.~(\ref{eq:Gp1}) is the only term in \Gp acting onto ${\cal A}_1$;
obviously it is suppressed for large $t$. Eq.~(\ref{eq:Gp2})
exemplifies a term in \Gp acting onto ${\cal A}_0$; it is independent
of $t$. This affirms the claim that, in the large $B$-field limit,
\Gp and \Gtp act only onto ${\cal A}_0$.

As a consequence, all BRST-like operators in the equations in~%
\cite{Lechtenfeld:2002cu} in the Seiberg-Witten limit act only onto the
oscillator algebra ${\cal A}_1$. Thus, if we assume a factorized
solution $\P=\P_0\otimes\P_1$ with $\P_0\in{\cal A}_0$ and $\P_1\in
{\cal A}_1$, the equation of motion can be restricted to ${\cal A}_0$
if $\P_1$ is chosen to be a projector (i.e., $\P_1\star\P_1=\P_1$):
\begin{equation}
  0 = \Gtp (\text{e}^{-\P} \Gp \text{e}^\P) = \Gtp (\text{e}^{-\P_0}
    \Gp \text{e}^{\P_0})\otimes \P_1 . \label{eq:SFTfact}
\end{equation}

Nevertheless, for finite $B$, the string field theory equation of motion
contains the ncSDYM equation of motion. Therefore, the solutions to be
constructed in the following section can serve as a guide in the search
for nonperturbative solutions to string field theory. Note that some
proposals for string functionals~$T$ were made in~\cite{Lechtenfeld:2002cu,
Kling:2002ht}; indeed, these solutions were motivated by the above ideas.

\section{Configurations without scattering} \label{sec:nointeract}
\noindent
The aim of this section is to demonstrate the usability of the solution
generating technique described in section~\ref{sec:dress} in two simple
cases. In 2+2 dimensions, we will construct an abelian GMS-like solution
of codimension four and a solution representing two nonabelian moving
lumps without scattering. The description of configurations with
scattering will be relegated to section~\ref{sec:scattering}. Although
we do not check their physical properties like tension and fluctuation
spectrum, we will refer to them as D-branes.

\subsection{Abelian GMS-like solution}
It is fairly easy to construct $U(1)$ solutions depending on all
space-time coordinates (i.e., with codimension four) via the dressing
approach in 2+2 dimensions (cf.~\cite{Horvath:2002bj} on euclidean
instantons via dressing). To this aim, let us start from the
discussion of the dressing approach in the hermitean gauge (section~%
\ref{sec:hermitean}). For $m=1$, we can omit all labels $k$; a
comparison of~(\ref{eq:newpsi})--(\ref{eq:Gtpkh}) with the
multiplicative ansatz shows that $\widetilde{R}=(|\mu'|^2-1)P$, where
$P$ is a hermitean projector independent of $\l$. If we choose $\th=\tht$
and define harmonic oscillators\footnote{Recall that $|\mu'|>1$ since
$\mu\in\mathbb{H}_-$ in section~\ref{sec:unitary}.}
\begin{equation}
  c_i := \frac{1}{\sqrt{2\th(1-|\mu'|^{-2})}}\eta^i \quad\text{and}\quad
  c_i^\dag := \frac{1}{\sqrt{2\th(1-|\mu'|^{-2})}} \eb^i,
\end{equation}
then $[c_i, c_j^\dag] = \de_{ij}$; thus, we can easily invert their
commutation relations and obtain
\begin{equation}
  \sqrt{2\th(1-|\mu'|^{-2})}\pa_{\eb^i} = \big[ c_i,. \big] .
    \label{eq:wbder}
\end{equation}
With this, we may rewrite~(\ref{eq:rkcond2}) as
\begin{subequations}
\begin{eqnarray}
  (1-P) c_1 P & = & 0 , \\
  (1-P) c_2 P & = & 0 .
\end{eqnarray}
\end{subequations}
Obviously, these equations can be solved by the projector $P=|0,0\>'\,
{}'\!\<0,0|$ onto the new vacuum $|0,0\>'$ annihilated by $c_1$ and
$c_2$.\footnote{Since the $a$-oscillators (cf.~(\ref{eq:wprdef}) and
(\ref{eq:aop})) and the $c$-oscillators are related by a unitary
transformation $c_i = U a_i U^\dag$, the (properly normalized) vacuum
$|0,0\>'$ can naturally be obtained as $|0,0\>'= U|0,0\>$.} We may use
the inverse Moyal-Weyl map~(\ref{eq:invMW}) to transform it to the star
formulation:
\begin{eqnarray}
  P_\star & = & \exp\left( -\frac{\eta^1\eb^1+\eta^2\eb^2}
    {\th(1-|\mu'|^{-2})} \right) \\
  & = & \exp\left( -\frac{(z^1+\mb'^{-1}\zb^2)(\zb^1+\mu'^{-1} z^2)
    +(z^2+\mb'^{-1}\zb^1)(\zb^2+\mu'^{-1} z^1)}
    {\th(1-|\mu'|^{-2})} \right). \nonumber
\end{eqnarray}
This is the analogue of the GMS-solution~\cite{Gopakumar:2000zd} in
2+2 dimensions; the projector $P_\star$ is an example for a projector
$\P_1$ in (\ref{eq:SFTfact}). The gauge potential can be derived
from~(\ref{eq:Yangg}) with $h^{-1} = 1-(1-|\mu'|^2) P_\star$. The
computation of the value of the action for this solution turns out to
be rather unwieldy.

\subsection[$U(2)$ solitons without scattering]{$\mathbf{U(2)}$ solitons
without scattering} \label{sec:solnoscat}
Let us now demonstrate how the additive ansatz~(\ref{eq:addansatz}) in
the unitary gauge can be employed to construct a solution describing
two moving lumps. A detailed description of the asymptotic space-time
picture will be given at the end of this section.

\noindent
{\bf Additive ansatz.} We work in the star formulation and relax the
condition that $\th=\tht$. The result of the first dressing step,
corresponding to a soliton in 2+1 dimensions, has already been given
in section~\ref{sec:dimredux}. This lump moves w.r.t.\ $t$ in the $xy$-%
plane; its energy (which is well-defined in 2+1 dimensions) was computed
in the same section. In the second dressing step, we want to add a soliton
confined for large $t$ to the $xy\tilde{t}$-plane. From the preceding
discussion in section~\ref{sec:unitary} it is clear that for $m=2$, we
can construct a solution to the self-duality equations using (cf.~%
(\ref{eq:addansatz}))
\begin{equation}
  \psi_2 = 1 + \sum_{l,k=1}^{2} \frac{T_l \Gamma^{lk}
    T_k^{\dag}}{\z-\mu_k} \label{eq:soladd}
\end{equation}
with $T_1=\left(\begin{smallmatrix} 1 \\ w_1 \end{smallmatrix}\right)$
and $T_2=\left(\begin{smallmatrix} 1 \\ \wt_2 \end{smallmatrix}\right)$.
However, it is not obvious that this solution really represents two
soliton-like objects, i.e., whether for large $t$ the solution can be
integrated over a (spatial) plane in the $xy\tilde{t}$-subspace to give
finite energy (and vice versa on a plane in the $xyt$-subspace at large
$\tilde{t}$). To prove this, we compare the additive (first-order pole)
and multiplicative ans\"atze for asymptotic times. Note that the two are
only equivalent if the multiplicative ansatz features merely first-order
poles in $\mu_i$, that is, if $\mu_1\neq\mu_2$.

\noindent
{\bf Multiplicative ansatz.} In the multiplicative ansatz, $\psi_2=
\chi_2\chi_1\psi_0$ may be constructed by two successive dressing steps
from a seed solution $\psi_0=1$. As in eq.~(\ref{eq:multansatz}), we
may write
\begin{equation}
  \psi_2 = \left( 1 +\frac{\mu_2 -\mb_2}{\z-\mu_2} P_2 \right)
    \left( 1 +\frac{\mu_1 -\mb_1}{\z-\mu_1}P_1 \right),
    \label{eq:solmult}
\end{equation}
and this has to coincide with~(\ref{eq:soladd}) for all times. Remember
that for hermitean projectors $P_k=\Tt_k(\Tt_k^\dag \Tt_k)^{-1}\Tt^\dag_k$
this ansatz guarantees the reality condition~(\ref{eq:gfreality1}).
The solution $\psi_2$ is subject to eqs.~(\ref{eq:laxrw}):
\begin{subequations} \label{eq:solLax}
\begin{eqnarray}
  \psi_2 (\z\pa_\vt+\pa_u ) \psi_2^{\dag} & = &  A_{2,u}, \\
  \psi_2 (\z\pa_v -\pa_\ut) \psi_2^{\dag} & = & -A_{2,\ut}.
\end{eqnarray}
\end{subequations}
The removability of the poles of the left hand sides at $\z=\mb_1$
and $\z=\mu_1$ is assured if (for $\mu_1\neq\mu_2$)
\begin{equation}
  (1-P_1) \bar{L}_1^1 P_1 = 0 \quad\text{and}\quad
  (1-P_1) \bar{L}_1^2 P_1 = 0 ,
\end{equation}
and this allows for a solution $\Tt_1=T_1=\left(\begin{smallmatrix}
1 \\ w_1 \end{smallmatrix}\right)$. Using the inverse Moyal-Weyl map,
we obtain for $P_1$ and its large-time limits
\begin{equation}
  P_{1\star} = \begin{pmatrix}
    \frac{2\th}{w\wb+\th} & \frac{\sqrt{2\th}w\wb^2}{(w\wb+\th)^2} \\
    \frac{\sqrt{2\th}^{\phantom{\dag}}w^2\wb}{(w\wb+\th)^2} &
    \frac{w\wb+\th}{w\wb+3\th}
  \end{pmatrix} \stackrel{t\to\pm\infty}{\longrightarrow}
  \begin{pmatrix} 1 & 0 \\ 0 & 0 \end{pmatrix} =:\Pi_{\pm\infty}
\end{equation}
with the ordinary product between $w_1$ and $\wb_1$, as in~%
(\ref{eq:onesol}).\footnote{From the asymptotics, we can read off
that the two lumps pass through each other without scattering.} In
contrast, $P_2$ will in general be a function $P_2(t,\wt_2,\ov{\wt}_2)$,
i.e., $\Tt_2\neq T_2$ may also depend on $t$. Namely, the residue
equation of~(\ref{eq:solLax}) at $\z=\mb_2$ and $\z=\mu_2$ yields:
\begin{subequations} \label{eq:solLaxmu2}
\begin{eqnarray}
  (1-P_2) \left( 1+\frac{\mu_1-\mb_1}{\mb_2-\mu_1}P_1 \right)\bar{L}_2^1
    \left\{ \left( 1+\frac{\mb_1-\mu_1}{\mb_2-\mb_1}P_1 \right)P_2
    \right\}= 0, \\
  (1-P_2) \left( 1+\frac{\mu_1-\mb_1}{\mb_2-\mu_1}P_1 \right)\bar{L}_2^2
    \left\{ \left( 1+\frac{\mb_1-\mu_1}{\mb_2-\mb_1}P_1 \right)P_2
    \right\}= 0.
\end{eqnarray}
\end{subequations}
Due to the asymptotic constancy of $P_1$ for large $|t|$, we can move
the Lax operators next to $P_2$ in this limit, and a short calculation
shows that this leads to
\begin{equation}
  (1-P_2)\pa_{\ov{\wt}_2} P_2 = 0 \quad\text{for }|t|\to\infty .
    \label{eq:solLaxP2}
\end{equation}
Obviously, we have $\Tt_2 = T_2$ only asymptotically.

Thus, the energy of the second lump can be computed in the limit
$|t|\to\infty$ to give $E_2=8\pi\cosh\eta_2\sin\varphi_2$ as in section~%
\ref{sec:dimredux}. Analogously, the energy of the first lump in the limit
$|\tilde{t}|\to\infty$ equals $E_1=8\pi\cosh\eta_1\sin\varphi_1$.

For large and fixed $|t|$, the space-time interpretation of the above
solution is as follows: Since $P_1$ is independent of $\tilde{t}$, the
first soliton (at a fixed time $t$) has some definite position in the
$xy$-plane and extends along the $\tilde{t}$-direction (see figure~1).
Moreover, the world-volume of the second soliton in this snapshot
corresponds to a tilted line (cf.\ eq.~(\ref{eq:solLaxP2})). When $t$
varies in the asymptotic region, the first (vertical) line gets shifted
in a direction determined by $\mu_1$, while the second line remains
fixed. Generically, the two world-volumes intersect the $xy$-plane at
different points. Since $\U$ depends on both $t$ and $\tilde{t}$, it is
not possible to perform a T-dualization in one of the time directions.
Thus, the solution has to be interpreted in terms of tilted D1-branes
inside space-time filling D-branes.
\begin{figure}[h]
\begin{center}
\resizebox{9.5cm}{7.3cm}{\includegraphics{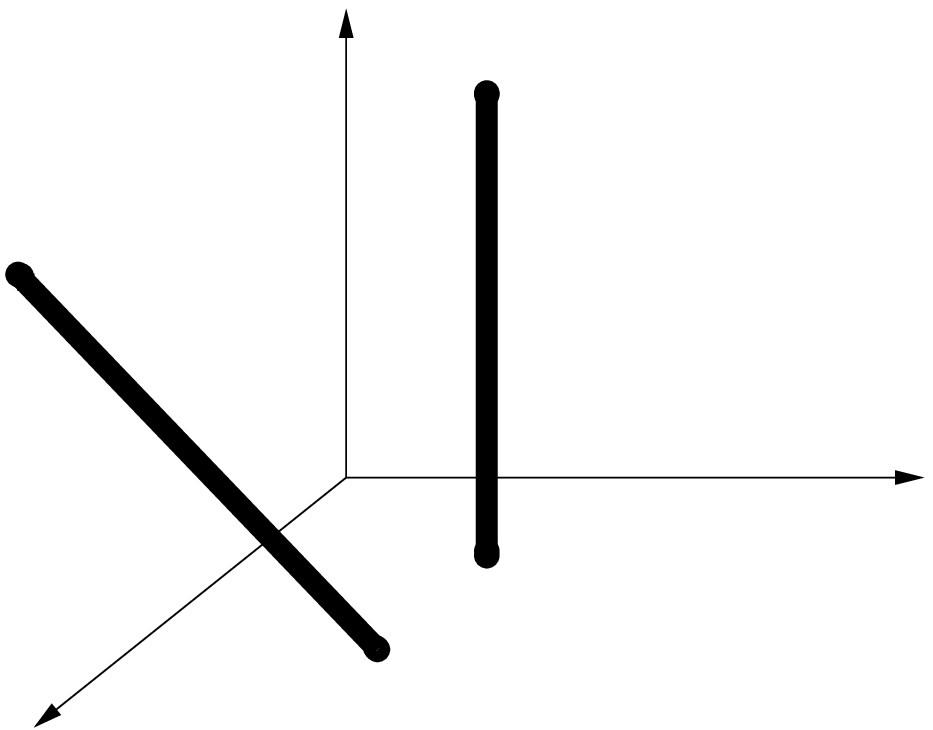}}
\setlength{\unitlength}{1cm}
\begin{picture}(0,0)
  \put(0.0,2.3){\makebox(0,0)[c]{$y$}}
  \put(-8.8,0.1){\makebox(0,0)[c]{$x$}}
  \put(-6.0,7.0){\makebox(0,0)[c]{$\tilde{t}$}}
  \put(-3.5,4.5){\makebox(0,0)[c]{soliton 1}}
  \put(-8.2,4.5){\makebox(0,0)[c]{soliton 2}}
\end{picture}
\end{center}
Figure 1:\quad Snapshot of the configuration discussed in section~%
\ref{sec:solnoscat} for fixed large $|t|$. The support of the solution
is concentrated around the solid lines.
\end{figure}

\section{Configurations with scattering}\label{sec:scattering}
\noindent
In this section we discuss two different setups entailing configurations
with scattering, namely two $U(2)$ soliton-like objects and
two noncommutative $U(2)$ plane waves, with world-volumes in 2+1
dimensional subspaces of $\R^{2,2}$. As will become clear in this section,
the crucial difference between the two configurations lies in the fact
that for these plane waves scattering occurs even if $\mu_1\neq\mu_2$,
whereas soliton-like objects only scatter nontrivially if $\mu_1=\mu_2$.
(In fact, we have already seen in the previous section that the solitonic
lumps do not scatter for $\mu_1\neq\mu_2$.)

\subsection[$U(2)$ solitons with scattering]{$\mathbf{U(2)}$ solitons
with scattering}\label{sec:sscat}
The setup is as in section~\ref{sec:solnoscat}; one of the solitonic
lumps is evolving with $t$ on the $xy$-plane. Since a first-order pole
ansatz for the auxiliary field $\psi_2$ in eqs.~(\ref{eq:soladd}) did
not lead to scattering, we now scrutinize the multiplicative ansatz with
$\mu_1=\mu_2=\mu$.

\noindent
{\bf First dressing step.} Starting from a seed solution $\psi_0=1$, we
make the following ansatz for the first dressing step:
\begin{equation}
  \psi_1 = 1 + \frac{\mu-\mb}{\z-\mu} P_1.
\end{equation}
This automatically fulfills the reality condition~(\ref{eq:gfreality1})
as long as $P_1$ is a hermitean projector. The residue condition on the
linear equations~(\ref{eq:laxrw}) leads to
\begin{equation}\label{eq:purew}
  (1-P_1) \pa_{\wb} P_1 = 0,
\end{equation}
i.e., $P_1$ varies in a $2+1$ dimensional subspace parametrized by $w$
as defined in~(\ref{eq:w_k}).

Now we set out to find explicit expressions for the components of the
gauge potential. First note that, since $P_1$ is chosen to be
independent of $\tilde{t}$, $A_{1, \ut}$ effectively reduces to
$A_{1,x}$. From eqs.~(\ref{eq:gfphi}) and using $\psi_2(\z=0)=\U^{-1}$
or~(\ref{eq:ominv}), we find\footnote{Alternatively, we could parametrize
$A_{1,u}$ and $A_{1,\ut}$ in terms of the algebra-valued Leznov
prepotential $\phi_1$ (cf.~\cite{Ivanova:2000zt}):
\begin{equation*}
  A_{1,\ut} = \pa_v\phi_1, \quad A_{1,u} = -\pa_\vt\phi_1,
\end{equation*}
where $\phi_1 = (\mu-\mb)P_1$ is defined by the asymptotic condition
\begin{equation*}
  \psi_1(u,v,\ut,\vt,\z\to\infty) = 1+ \z^{-1}
    \phi_1(u,v,\ut,\vt) + {\mathcal{O}}(\z^{-2}).
\end{equation*}
This also leads to eqs.~(\ref{eq:scatsol}).}
\begin{subequations}\label{eq:yang}
\begin{eqnarray}
  A_{1,u} = \bar{\r}\left( 1-\r P_1\right) \pa_u P_1, \label{eq:yang1} \\
  A_{1,x} = \bar{\r}\left( 1-\r P_1\right) \pa_x P_1, \label{eq:yang2}
\end{eqnarray}
\end{subequations}
where $\r=1-\mb/\mu$ was introduced for convenience.

\noindent
{\bf Second dressing step.} The reality condition~(\ref{eq:gfreality1})
for the new ansatz $\psi_2=\chi_2\psi_1$ will be satisfied if we choose
$\chi_2$ to be of the same functional form as $\psi_1$, i.e.,
\begin{equation}
  \psi_2 = \left( 1+\frac{\mu-\mb}{\z-\mu}P_2\right) \left( 1+
    \frac{\mu-\mb}{\z-\mu} P_1\right)
\end{equation}
with a hermitean projector $P_2 = T_2(T_2^\dag T_2)^{-1}T_2^\dag$ in
general depending on all four coordinates. The corresponding gauge-fixed
linear equations~(\ref{eq:laxrw}) are:
\begin{subequations}
\begin{eqnarray}
  \psi_2 (\z\pa_\vt+\pa_u) \psi_2^{\dag} = A_{2,u}, \\
  \psi_2 (\z\pa_v-\pa_\ut) \psi_2^{\dag} = -A_{2,\ut} ,
\end{eqnarray}
\end{subequations}
which is equivalent to
\begin{subequations}
\begin{eqnarray}
  A_{2,u} & = & \chi_2 A_{1, u} \chi_2^{\dag} + \chi_2 (\z\pa_\vt
    +\pa_u) \chi_2^{\dag}, \label{eq:secdress1}\\
  -A_{2,\ut} & = & \chi_2 A_{1,\ut} \chi_2^{\dag}
    -\chi_2 (\z\pa_v-\pa_\ut) \chi_2^{\dag}. \label{eq:secdress2}
\end{eqnarray}
\end{subequations}
Inserting $\chi_2 = 1+\frac{\mu-\mb}{\z-\mu}P_2$ and demanding that the
right hand sides of eqs.~(\ref{eq:secdress1}) and (\ref{eq:secdress2})
are free of poles for $\z\to\mb$ and $\z\to\mu$ leads to
\begin{subequations}\label{eq:master}
\begin{gather}
  (1-P_2) \left\{ \r\pa_{\wb^1}-A_{1,u}\right\} P_2 = 0 ,
    \label{eq:master1} \\
  (1-P_2) \left\{ \r\pa_{\wb^2}+A_{1,x}\right\} P_2 = 0.
    \label{eq:master2}
\end{gather}
\end{subequations}
Recall that we defined $\r=1-\mb/\mu$. In the following, we shall assume
$P_2 = P_2(w,\wb,\tilde{t})$. By appropriately combining eqs.~%
(\ref{eq:master}) and taking into account eqs.~(\ref{eq:yang}), we obtain
the following equations for the projector $P_2$:
\begin{subequations}\label{eq:scatsol}
\begin{align}
  & (1-P_2) \left\{ \pa_{\wb}P_2 -(\pa_{\wb}P_1) P_2 \right\} = 0, \\
  & (1-P_2) \left\{ \pa_{\tilde{t}}P_2+\nu\bar{\r}(\pa_w P_1)
    P_2 \right\} = 0.
\end{align}
\end{subequations}
In the derivation of the second equation we have also
made use of the hermitean conjugate of eq.~(\ref{eq:purew}), that is
\begin{equation}
  P_1 \pa_w P_1 = 0.
\end{equation}

In the operator formalism, all derivatives can be understood as
commutators in the sense of~(\ref{eq:opcdef}). The projector
identities
\begin{equation}\label{eq:projid}
  (1-P_2)P_2\equiv 0 \quad\text{and}\quad (1-P_2)T_2\equiv 0
\end{equation}
transform eqs.~(\ref{eq:scatsol}) into
\begin{subequations}\label{eq:opeq1}
\begin{align}
  & (1-P_2)\big\{ w T_2-[w,P_1]T_2 \big\} (T_2^{\dag}T_2)^{-1}
    T_2^{\dag} = 0, \\
  & (1-P_2)\big\{ \pa_{\tilde{t}} T_2 - \eta' [\wb, P_1] T_2
    \big\} (T_2^{\dag}T_2)^{-1} T_2^{\dag} = 0,
\end{align}
\end{subequations}
where $\eta':=\frac{\nu}{2\th}\mb^{-1}(\mb-\mu)$ and $\nu=\nu_1=
\nu_2$ from~(\ref{eq:nu}). Due to (\ref{eq:projid}), a sufficient
condition for a solution is given by
\begin{subequations}\label{eq:scat1}
\begin{eqnarray}
  w T_2-[w,P_1]T_2 = T_2{\mathcal{S}}_1 \label{eq:scat1a}\\
  \pa_{\tilde{t}}T_2-\eta' [\wb,P_1]T_2 = T_2{\mathcal{S}}_2,
    \label{eq:scat1b}
\end{eqnarray}
\end{subequations}
for some functions ${\mathcal{S}}_1(w,\wb,\tilde{t})$ and
${\mathcal{S}}_2(w,\wb,\tilde{t})$.

\noindent
{\bf Explicit solutions.} For the example of $U(2)$ soliton-like
configurations, we choose $T_1=\left(\begin{smallmatrix} 1 \\ w
\end{smallmatrix}\right)$, which is the simplest nontrivial $U(2)$
ansatz compatible with eq.~(\ref{eq:purew}). In the operator formalism,
\begin{equation}
  P_1 = T_1 (T_1^{\dag}T_1)^{-1} T_1^{\dag} = \begin{pmatrix}
    (1+ \wb w)^{-1} & (1+ \wb w)^{-1} \wb \\
    w(1+ \wb w)^{-1} & w(1+ \wb w)^{-1}\wb \end{pmatrix}.
\end{equation}
Our task is now to determine a possible solution for $T_2$. We employ
the ansatz
\begin{equation}\label{eq:t2ansatz}
  T_2 = \begin{pmatrix} u_1(\tilde{t},w,\wb) \\ u_2(\tilde{t},w,
    \wb) \end{pmatrix} .
\end{equation}
Setting ${\mathcal{S}}_1=w$ and inserting~(\ref{eq:t2ansatz}) into
eq.~(\ref{eq:scat1a}) yields
\begin{subequations}
\begin{eqnarray}
  \big[ w,u_1 \big] & = & \big[ w,(1+\wb w)^{-1} \big] (u_1 +
    \wb u_2) + 2 \th(1+\wb w)^{-1} u_2, \label{eq:addcond} \\
  \big[ w,u_2 \big] & = & w\big[ w,(1+\wb w)^{-1} \big] (u_1 +
    \wb u_2) + 2 \th w (1+\wb w)^{-1} u_2 .
\end{eqnarray}
\end{subequations}
The last two equations immediately imply
\begin{equation}\label{eq:scat2a}
  [ w, w u_1 - u_2 ] = 0.
\end{equation}

\noindent
{\bf The case $\pmb{\tht = 0}$.} Evidently, if we restrict ourselves to
$[t,\tilde{t}]=\ic\tht = 0$,
\begin{equation}
  u_2 = w u_1 - f(\tilde{t}, w)
\end{equation}
solves eq.~(\ref{eq:scat2a}) with an arbitrary function $f$ (depending
only on $\tilde{t}$ and $w$) yet to be determined.
Exploiting eqs.~(\ref{eq:addcond}) and~(\ref{eq:scat2a}), we find a
solution
\begin{equation}\label{eq:uansatz}
  u_1 = 1 + (1+\wb w)^{-1}\wb f(\tilde{t},w), \quad
  u_2 = w - (1+\wb w + 2\th)^{-1} f(\tilde{t},w).
\end{equation}
From~(\ref{eq:scat1b}) we obtain in a similar fashion
\begin{equation}\label{eq:scat2b}
  \pa_{\tilde{t}} u_1 = \eta'\big[ \wb,(1+\wb w)^{-1} \big]
    (u_1 + \wb u_2)
\end{equation}
by setting ${\mathcal{S}}_2= 0$. Taking into account eq.~(\ref{eq:scat2b})
the explicit $\tilde{t}$-dependence of $f(\tilde{t},w)$ can be easily
deduced:
\begin{equation}
  f = 2\th\eta'\left( \tilde{t}+h(w) \right),
\end{equation}
for some function $h$ meromorphic in $w$. Finally, substituting the
results into~(\ref{eq:t2ansatz}) leads to
\begin{equation}
  T_2 = \begin{pmatrix} 1 \\ w \end{pmatrix} + \begin{pmatrix}
    \wb \\ -1 \end{pmatrix} (1+\wb w+2\th)^{-1} f(\tilde{t},w).
\end{equation}
Translating this to the star formalism, we easily read off that $T_1=T_2$
at the zero locus of $f_\star(\tilde{t},w)$; moreover, if we restrict $\mu$
to be purely imaginary, $\mu=-\ic p$, $p\in(1,\infty)$, $\U$ degenerates at
these points to the identity. If we choose $h_\star = w \star w = w^2$,
which corresponds to two moving soliton-like objects, this leads to right
angle scattering~\cite{Lechtenfeld:2001gf}:
\begin{equation}
  f_\star = 0 \Rightarrow w = \pm \sqrt{-\tilde{t}}. \label{eq:locus}
\end{equation}
For the points in this locus, $w$ is purely real for $\tilde{t}<0$, and
$w$ is purely imaginary for $\tilde{t}>0$. Since for the above choice of
$\mu$,
\begin{equation}
  w = \left( \frac{2}{p+p^{-1}}\right)^{1/2} \left( x+\frac{\ic}{2}
    (p+p^{-1})y+\frac{\ic}{2}(p-p^{-1})t \right), \label{eq:wmuimag}
\end{equation}
we see that e.g.\ for $t=0$, the point where $\U=1$ moves along the
positive $x$-axis accelerating towards the origin for negative
$\tilde{t}$. For positive $\tilde{t}$ it decelerates during its motion
along the positive (or negative, depending on the sign in~%
(\ref{eq:locus})) $y$-axis.

\noindent
{\bf The case $\pmb{\tht\neq 0}$.}
If the two time directions do not mutually commute, i.e., $\tht\neq 0$,
eq.~(\ref{eq:scat1b}) can be written as
\begin{equation}\label{eq:scat1c}
  \frac{1}{\ic\tht} [t,T_2]-\eta'[w,P_1] T_2 = T_2{\mathcal{S}}_2.
\end{equation}
Now, we can still solve eq.~(\ref{eq:scat2a}) by
\begin{equation}
  u_2 = w u_1 - g(\tilde{t},\wb,w).
\end{equation}
The difference to the case $\tht=0$ is that now the vanishing of the
commutator~(\ref{eq:scat2a}) can only be achieved by a nontrivial
choice for $g(\tilde{t},\wb,w)$, e.g.,
\begin{equation}
  g(\tilde{t},\wb,w) = \tilde{t} + \a\wb + h(w),
\end{equation}
where $\a:=-\frac{\ic}{4}(\mb + \mb^{-1})\frac{\tht}{\th}$
and $h(w)$ is again an arbitrary function meromorphic in $w$. Let us
restrict $\mu$ again to be purely imaginary, $\mu=-\ic p$, $p\in(1,
\infty)$, then $\a\in\R_+$.

Apparently we also need $\th\neq 0$; then, the contributions of $[w,\wb]$
and $[w,\tilde{t}]$ add up to zero. If we use the inverse Moyal-Weyl map
to translate to the star product and choose $h_\star (w)= w\star w= w^2$,
we obtain
\begin{equation}
  g_\star(\tilde{t},\wb,w)= \tilde{t} + \a\wb + w^2.
\end{equation}
The subsequent calculation is analogous to the case $\tht=0$. It turns
out that $P_1$ and $P_2$ coincide and $\U=1$ at the locus of
$g_\star(\tilde{t},\wb,w)$, i.e., $\tilde{t} + \a\wb + w^2 =0$. If we
split $w$ into real and imaginary parts,
\begin{equation}
  w = a+\ic b,
\end{equation}
we can easily read off $a$ and $b$ from eq.~(\ref{eq:wmuimag}), and
the locus where $\U=1$ is given by
\begin{equation}
  -\tilde{t} = \a a+a^2-b^2 + \ic(2a-\a)b .
\end{equation}
Since $\tilde{t}$ is real, obviously either $b=0$ or $a=\a/2$. We
obtain
\begin{subequations}
\begin{align}
  b & = 0 & \!\!\!\!\Longrightarrow \qquad a & = -\frac{\a}{2}\pm
    \sqrt{\frac{\a^2}{4}-\tilde{t}} & & \!\!\!\!\mbox{for}\qquad
    \tilde{t}\leq\frac{\a^2}{4}, \\
  a & = \frac{\a}{2} & \!\!\!\!\Longrightarrow \qquad b & = \pm
    \sqrt{\frac{3}{4}\a^2+\tilde{t}} & & \!\!\!\!\mbox{for}\qquad
    \tilde{t}\geq -\frac{3}{4}\a^2 .
\end{align}
\end{subequations}

\begin{figure}[h]
\begin{center}
\resizebox{82mm}{5cm}{\includegraphics{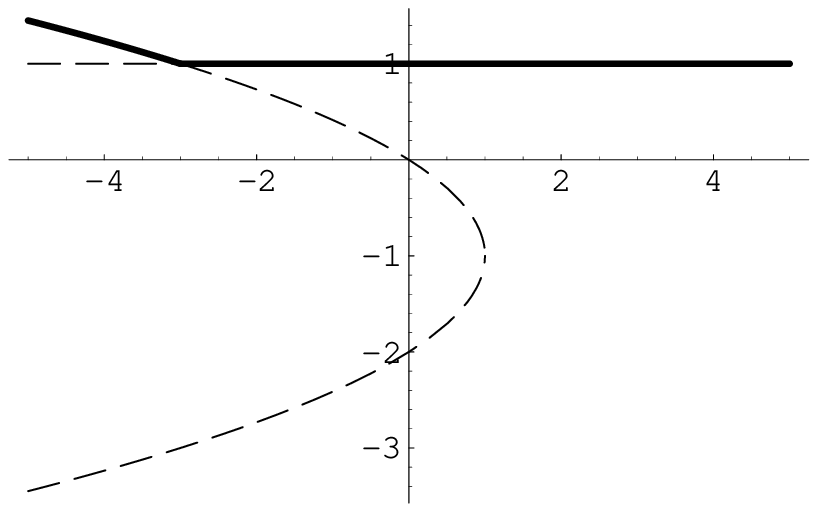}}
\setlength{\unitlength}{1cm}
\begin{picture}(0,0)
\put(0.0,3.5){\makebox(0,0)[c]{$\tilde{t}$}}
\put(-6.7,3.0){\makebox(0,0)[c]{$-\frac{3}{4}\a^2$}}
\put(-1.5,4.7){\makebox(0,0)[c]{$a(\tilde{t})=\frac{\a}{2}$}}
\put(-2.5,1.2){\makebox(0,0)[c]{$a(\tilde{t})=-\frac{\a}{2}\pm
  \sqrt{\frac{\a^2}{4}-\tilde{t}}$}}
\thinlines
\put(-6.67,3.35){\line(0,1){0.2}}
\end{picture} \\[2mm]
\resizebox{82mm}{5cm}{\includegraphics{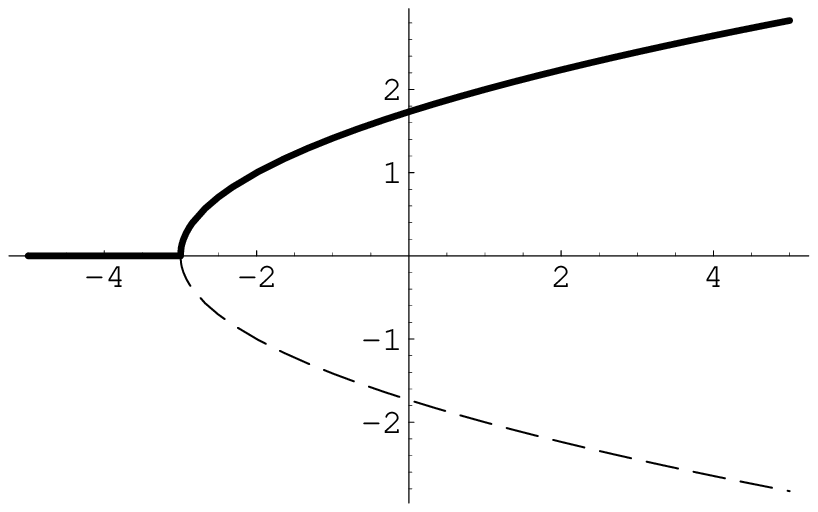}}
\setlength{\unitlength}{1cm}
\begin{picture}(0,0)
\put(0.0,2.5){\makebox(0,0)[c]{$\tilde{t}$}}
\put(-8.0,2.8){\makebox(0,0)[c]{$b(\tilde{t})=0$}}
\put(-2.5,1.2){\makebox(0,0)[c]{$b(\tilde{t})=\pm\sqrt{\frac{3}{4}\a^2
  +\tilde{t}}$}}
\end{picture}
\end{center}
Figure 2:\quad Motion of the ``point of degeneracy'' where $\U=1$ in
$\tilde{t}$ (bold lines). Its coordinates $a$ and $b$ are plotted for
$\a=2$. For $b(\tilde{t})$, exemplarily the upper branch was chosen.
\end{figure}

This can be interpreted as follows: The ``point of degeneracy'' where
$\U=1$ moves along $a=-\a/2+\sqrt{\a^2/4-\tilde{t}}$ and $b=0$ as
$\tilde{t}$ grows until $\tilde{t}=-3\a^2/4$. Then, as $\tilde{t}$
grows larger, it moves along $a=\a/2$ and $b=\pm\sqrt{3\a^2/4+
\tilde{t}}$ (see figure~2). With the help of~(\ref{eq:wmuimag}), it is
easy to interpret this motion in the $xy$ plane (for fixed $t$). Therefore
we have shown that it is possible to construct nontrivial configurations
with scattering also for the case of noncommuting time directions.
More complicated solutions in both cases may be constructed by making
different choices for $h(w)$ or by choosing a more sophisticated ansatz
for $T_1$ and $T_2$.

\subsection{Colliding plane waves}\label{sec:pwscat}
\noindent
Beside the soliton-like solutions (discussed above), there is another
class of exact solutions to the self-duality equations~(\ref{eq:realsd}),
namely extended plane waves. For asymptotic times, each of them has
codimension two. In the commutative case, these were constructed and
discussed in~\cite{deVega:1987dj, Korepin:1996mm, Rosly:1996vr}. In the
context of the $U(N)$ sigma model in 2+1 dimensions, this type of
solution was first discussed by Leese~\cite{Leese:hj}; the noncommutative
generalization was given in~\cite{Bieling:2002is}. In~\cite{Gross:2000ss},
plane waves were described in (noncommutative) D1-D3 systems. Here we
want to show that one can construct noncommutative two-wave solutions in
ncSDYM which entail nontrivial scattering even for $\mu_1\neq\mu_2$.

\noindent
{\bf Additive ansatz.} We assume $\mu_1\neq\mu_2$ henceforth and therefore
make a single-pole ansatz for the auxiliary field~$\psi$. In this section,
exceptionally all products are understood to be star products (including
the inverse and the exponential of coordinates). The calculation is
largely parallel to the derivation in section~\ref{sec:solnoscat} which
gives us the opportunity to shorten the description here and to
concentrate on the novel features.

We start from the additive ansatz~(\ref{eq:soladd}), but now choose,
inspired by~\cite{Leese:hj, Bieling:2002is}, the exponential ans\"atze
\begin{equation}\label{eq:exp1ansatz}
  T_1 = \begin{pmatrix} 1 \\ {\text e}^{b_1 w_1} \end{pmatrix}
    \quad\text{and}\quad T_2 = \begin{pmatrix} 1 \\ {\text e}^{b_2
    \wt_2} \end{pmatrix}
\end{equation}
with $b_1 \in\R_{>0}, b_2\in\R$. The discussion in section~%
\ref{sec:unitary} guarantees that this will yield a solution to the
self-duality equations. However, it is not obvious that this solution
factorizes into two plane waves for asymptotic times; to prove this,
we have to compare with the multiplicative ansatz again.

\noindent
{\bf Multiplicative ansatz.} The multiplicative ansatz takes the same
form as eq.~(\ref{eq:solmult}). It can be easily shown as in section~%
\ref{sec:solnoscat} that $P_1=\Tt_1(\Tt_1^\dag\Tt_1)^{-1}\Tt_1^\dag$
can consistently be constructed from $\Tt_1 = T_1$, given in~%
(\ref{eq:exp1ansatz}). Let us now scrutinize the $|t|\to\infty$ limits
of $P_1$. For simplicity, we set $\mu_1 = \ic p$ strictly imaginary with
$p>1$. Therefore, $\b_1$ in~(\ref{eq:w}) is real and
\begin{equation}
  \b_1 = - \frac{1}{2} \th^{-1/2} (p-p^{-1})(p+p^{-1})^{-1/2}< 0.
\end{equation}
If we consider the large $t$ limit, it turns out that $w_1$ is dominated
by the term linear in $t$, namely:
\begin{equation}
  b_1 w_1 \simeq\pm b_1\sqrt{2\th}|\b_1|t, \quad\text{for }t\to\pm\infty.
\end{equation}
Thus, $P_1$ in the large $t$ limit behaves as\\
\parbox{10.4cm}{\begin{equation*}
  P_1 = \begin{pmatrix} (1+\text{e}^{b_1\wb_1}
    \text{e}^{b_1 w_1})^{-1}
    & (1+\text{e}^{b_1\wb_1} \text{e}^{b_1 w_1})^{-1}\text{e}^{b_1\wb_1}
    \\ \text{e}^{b_1 w_1} (1+\text{e}^{b_1\wb_1} \text{e}^{b_1 w_1})^{-1^%
    {\phantom{\dag}}}
    & \text{e}^{b_1 w_1} (1+\text{e}^{b_1\wb_1} \text{e}^{b_1 w_1})^{-1}
    \text{e}^{b_1\wb_1} \end{pmatrix} \left\{\mbox{\rule[-1cm]{0cm}{2cm}}
    \right.
\end{equation*}}
\parbox{6.0cm}{
\begin{subequations}\label{eq:constproj}
\begin{align}
  \stackrel{t\to +\infty}{\longrightarrow}
  & \begin{pmatrix} 0 & 0 \\ 0 & 1 \end{pmatrix} =:\Pi_{+\infty}, \\
  \stackrel{t\to -\infty}{\longrightarrow}
  & \begin{pmatrix} 1 & 0 \\ 0 & 0 \end{pmatrix} =:\Pi_{-\infty}.
\end{align}
\end{subequations}}\\
In these limits, $P_1$ obviously becomes a constant projector. Again, the
Lax operators in~(\ref{eq:solLaxmu2}) can be moved next to $P_2$ in these
limits to give~(\ref{eq:solLaxP2}). This concludes the proof that we may
write $\Tt_2=T_2$ asymptotically.

In addition, we can conclude that this setup entails nontrivial
scattering, again by analyzing $\U^{\dag}=\psi_2(\z=0)$ in the limits
$t\to\pm\infty$:
\begin{subequations}
\begin{align}
  \U^\dag \Big{|}_{t \to +\infty} = \lim_{t\to +\infty}\psi_2(\z=0) =
    (1-\r_2 P_2)(1-\r_1\Pi_{+\infty}), \\
  \U^\dag \Big{|}_{t\to -\infty} = \lim_{t\to -\infty}\psi_2(\z=0) =
    (1-\r_2 P_2)(1-\r_1\Pi_{-\infty}).
\end{align}
\end{subequations}
For convenience, we have set $\r_k=1-\mb_k/\mu_k$ for $k=1,2$.
Clearly, $\U^{\dag}\Big{|}_{t\to +\infty}$ and $\U^{\dag}\Big{|}_{t\to
 -\infty}$ are different, which indicates nontrivial scattering behavior.

If we now additionally take $\tilde{t}\to\pm\infty$, we find that $P_2$
also becomes a constant projector,
\begin{equation}
  \lim_{\tilde{t}\to\pm\infty}P_2 = \Pi_{\pm\infty}.
\end{equation}
Therefore,
\begin{subequations}
\begin{align}
  \U^{\dag}\Big{|}_{t,\tilde{t}\to -\infty} & = \begin{pmatrix}
    \g & 0 \\ 0 & 1 \end{pmatrix}, \\
  \intertext{and}
  \U^{\dag}\Big{|}_{t,\tilde{t}\to +\infty} & = \begin{pmatrix}
    1 & 0 \\ 0 & \g \end{pmatrix},
\end{align}
\end{subequations}
where $\g:=\mb_1\mb_2\mu_1^{-1}\mu_2^{-1}$. Again, this result shows
the existence of scattering in this two-wave configuration.

The above-described solutions represent 1+2 dimensional plane waves
in the asymptotic domain, i.e., long before and after the interaction.
This can be seen by analyzing the energy density in 2+1 dimensional
subspaces, e.g., the energy density for a gauge field constructed
from $P_1$ (at a fixed time $\tilde{t}$) turns out to depend only on
one spatial direction~\cite{Bieling:2002is}. The asymptotic space-time
interpretation for this setup can be visualized by the following
snapshot for fixed large $t$ (see figure~3). Since $P_1$ is independent
of $\tilde{t}$, the corresponding wave extends along this direction.
The above energy density argument explains its spatial extension.
Observe that for this type of solutions, the moduli~$\mu_1$, $\mu_2$
not only parametrize the velocities of the plane waves but also their
respective parallel directions in the $xy$-plane~(cf.\ eqs.~%
(\ref{eq:w_k}) and~(\ref{eq:wt_k}) together with~(\ref{eq:exp1ansatz})).
When $t$ varies in the asymptotic region, the world-volume of the
first plane wave undergoes a parallel shift. Consider a space-like
section (i.e., $t$ {\sl and} $\tilde{t}$ fixed). Then, the intersection
of the two plane waves with this $xy$-plane will consist of two lines
which generically include some angle determined by the moduli $\mu_1$
and $\mu_2$. For later times $t$ or $\tilde{t}$, the lines corresponding
to $P_1$ and $P_2$ have changed position in the $xy$-subspace but
kept their directions.
\begin{figure}[h]
\begin{center}
\resizebox{9.0cm}{8.0cm}{\includegraphics{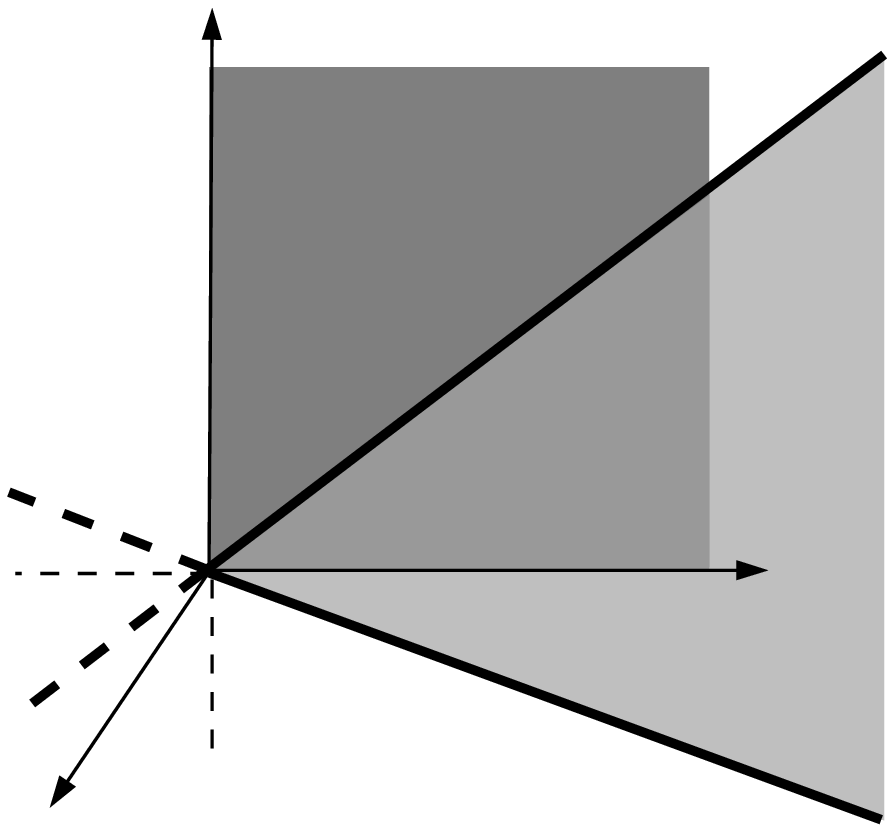}}
\setlength{\unitlength}{1cm}
\begin{picture}(0,0)
  \put(-1.3,2.5){\makebox(0,0)[c]{$y$}}
  \put(-8.4,0.1){\makebox(0,0)[c]{$x$}}
  \put(-6.8,8.0){\makebox(0,0)[c]{$\tilde{t}$}}
  \put(-5.0,6.6){\makebox(0,0)[c]{plane wave 1}}
  \put(-2.7,1.7){\makebox(0,0)[c]{plane wave 2}}
\end{picture}
\end{center}
Figure~3:\quad Snapshot for fixed large $|t|$ of one octant of the
configuration discussed in section~\ref{sec:pwscat}. For simplicity,
the first plane wave was chosen to be static ($\mu_1=-\ic$). This
choice implies that the energy density for the first wave at fixed
$\tilde{t}$ does not depend on $y$. The support of the solution is
concentrated around the grey planes.
\end{figure}

\section{Conclusions}
\noindent
In this note we have discussed exact solutions to the
self-duality equations of noncommutative Yang-Mills theory
on ${\mathbb R}^{2,2}$. To this aim, a Lax pair has been gauged in two
inequivalent ways; appropriate ans\"atze for the auxiliary field $\psi$
have been discussed. From concrete solutions $\psi$ to the residue
equations of the Lax pair explicit expressions for the gauge potentials
have been constructed. We have shown that the Lax pair is included in the
string field theoretic one; therefore, it seems plausible that this also
applies to its solutions. Conversely, our field theoretic solutions
could serve as a guideline to construct nonperturbative solutions of
N=2 string field theory. It seems reasonable to expect that a similar
program could be carried out for N=1 strings.

A GMS-like solution and solutions describing $U(2)$ solitons have been
constructed. Moreover, it has been shown that dimensional reduction to
2+1 dimensions leads to results coinciding with those of~%
\cite{Lechtenfeld:2001aw}--\cite{Wolf:2002jw}. Explicitly, the field
theory description of D-brane scattering (for plane wave and soliton-%
like configurations) has been generalized to the 2+2 dimensional case. It
would be interesting to trace this description to the string theory level,
i.e., compute scattering in the given $B$-field background by closed
string exchange. To corroborate the interpretation of our field theory
solutions as lower-dimensional D-branes one could try to compute their
Chern characters and examine the fluctuation spectrum around these
solutions.

\section*{Acknowledgments}
\noindent
The authors are grateful to O. Lechtenfeld for useful comments.
They also want to express their gratitude to A.D. Popov for many
helpful discussions on this subject.
This work was done within the framework of the DFG priority program
(SPP 1096) in string theory.

\begin{appendix}

\section{Self-duality, twistor space and holomorphicity} \label{sec:twistor}
\noindent
In this section, we explain the geometric setup underlying the method
we use to solve the self-duality equations of Yang-Mills theory on
$\R^{2,2}$. We mostly restrict ourselves to the commutative case,
comments on the noncommutative generalization are added where
appropriate.\footnote{For a description of twistors in the noncommutative
case, see~\cite{Kapustin:2000ek, Takasaki:2000vs, Lechtenfeld:2001ie}.}
For our purposes, $U(N)$ Yang-Mills theory is formulated in terms of a
$GL(N,\C)$ principal bundle $P\cong\R^{2,2}\times GL(N,\C)$ over the
(pseudo-)Riemannian ``space-time'' manifold $\R^{2,2}$. This principal
bundle should be endowed with an irreducible $GL(N,\C)$ connection~$A$
and its respective curvature $F$. We will impose a reality condition on
$A$ below. The self-duality equations $F=\ast F$ are tackled with the
help of a Lax pair, whose geometrical meaning will now be described.

\subsection{Isotropic coordinates}
\noindent
We will see in section~\ref{sec:tw} that the self-duality equations on
$\R^{2,2}$ can be written in real coordinates $x^\mu$ as
\begin{equation}
  \ov{W}_1^\mu \ov{W}_2^\nu F_{\mu\nu} = 0 \label{eq:sdw}
\end{equation}
for certain 4-vectors $\ov{W}_i$. To derive constraints on the $\ov{W}_i$,
it turns out to be useful to switch to a spinor notation. Exploiting
that $so(2,2)\cong sl(2,\R)\times sl(2,\R)$, we can rewrite $\ov{W}_i$ as
\begin{equation}
  (\ov{W}_i^{\ad\a}) = (\ov{\t}_\mu^{\ad\a} \ov{W}_i^\mu) =
  \begin{pmatrix} \ov{W}_i^4+\ov{W}_i^2 & \ov{W}_i^1 - \ov{W}_i^3 \\
    \ov{W}_i^1 + \ov{W}_i^3 & \ov{W}_i^4 - \ov{W}_i^2 \end{pmatrix}
  \label{eq:realsp}
\end{equation}
with the help of $SL(2,\R)$-generators~$\ov{\t}_a$, $a=1,2,3$ and
$\ov{\t}_4=1$. If we define as for the Pauli matrices $\t^\mu_{\b\bd}
=\eta^{\mu\nu}\ov{\t}_\mu^{\ad\a}\ve_{\ad\bd}\ve_{\a\b}$ with
$\ve_{12}=-1$, eq.~(\ref{eq:sdw}) can be rewritten as
\begin{equation}
  \ov{W}_1^{\ad\a} \ov{W}_2^{\bd\b} (F_{\a\b}\ve_{\ad\bd} +
    F_{\ad\bd}\ve_{\a\b}) = 0 .
\end{equation}
These are the self-duality equations $F_{\ad\bd}=0$ iff we choose
\begin{equation}
  \ov{W}_1^{\ad\a}=\xi^\a \pi^\ad \quad\text{and}\quad
    \ov{W}_2^{\bd\b}=\chi^\b \pi^\bd , \label{eq:iso}
\end{equation}
with arbitrary commutative spinors $\xi^\a, \chi^\a,$ and $\pi^\ad$.
That is, $\ov{W}_1$ and $\ov{W}_2$ have to span a null plane in
$\R^{2,2}$.

On $\R^{2,2}$, there are two possibilities to satisfy~(\ref{eq:iso}),
related to the existence of Majorana-Weyl spinors in 2+2 dimensions:
One can choose complex or real spinors. Since the $\ov{\t}$-matrices
in~(\ref{eq:realsp}) are real, this will lead to complex and real
coordinates on $\R^{2,2}$.

\subsection{Complex coordinates}
\noindent
{\bf Almost complex structures on $\R^{2,2}$.}
To elucidate the meaning of the $\ov{W}_i$ it is necessary to introduce
an {\sl almost complex structure} on $\R^{2,2}$. An almost complex
structure is a tensor field $J$ of type~(1,1) such that $J_\mu{}^\nu
J_\nu{}^\l=-\de_\mu{}^\l$. We shall consider translationally invariant
(constant) and therefore integrable almost complex structures, i.e.,
complex structures. It is easy to see that complex structures on
$\R^{2,2}$ are parametrized by the coset $SO(2,2)/U(1,1)\cong
SO(2,1)/SO(2)$.\footnote{For a given almost complex structure $J$
one can choose coordinates x$^1$, x$^2$, y$^1$, y$^2$ such that in this
basis, $J$ as an endomorphism of the tangent bundle maps $J(\tfrac{\pa}
{\pa\text{x}^k})=\tfrac{\pa}{\pa\text{y}^k}$ and $J(\tfrac{\pa}{\pa
\text{y}^k}) = -\tfrac{\pa}{\pa\text{x}^k}$ for $k=1,2$. A linear
combination $\pa_{\text{x}^k}-\ic J\pa_{\text{y}^k}=:\pa_{z^k}$ (as a
section of the complexified tangent bundle to $\R^{2,2}$) obviously
has eigenvalue $\ic$, it only rotates homogeneously under rotations
$M^n_k$ of the structure group $SO(2,2)$ if $J$ and $M$ commute. This
singles out a subgroup $U(1,1)$ of $SO(2,2)$ which leaves the fixed
complex structure invariant.} Without loss of generality, we can
restrict the discussion to almost complex structures compatible with
the metric (so that the metric is hermitean). Then, (anti)holomorphic
basis vectors are automatically null vectors.

One can realize~\cite{Helgason} this coset space on $so(2,1)$ in the
following way~\cite{Ivanova:rc}: We start from a matrix representation
of $so(2,1)$,
\begin{gather}
  I_1 = \begin{pmatrix} 0 & 0 & 0 & 1 \\ 0 & 0 & 1 & 0 \\
                        0 & 1 & 0 & 0 \\ 1 & 0 & 0 & 0 \end{pmatrix},
                        \qquad
  I_2 = \begin{pmatrix} 0 & 0 & -1 & 0 \\ 0 & 0 & 0 & 1 \\
                        -1 & 0 & 0 & 0 \\ 0 & 1 & 0 & 0 \end{pmatrix},
                        \nonumber \\
  I_3 = \begin{pmatrix} 0 & 1 & 0 & 0 \\ -1 & 0 & 0 & 0 \\
                        0 & 0 & 0 & 1 \\ 0 & 0 & -1 & 0 \end{pmatrix},
\end{gather}
satisfying $I_a I_b = g_{ab} + f_{ab}{}^c I_c$ with structure constants
$f_{12}{}^3=-f_{23}{}^1=-f_{31}{}^2=1$ and metric $(g_{ab})=\text{diag}
(1,1,-1)$ on $so(2,1)$. Then we can write a general complex structure
in the form
\begin{equation}
  J = -s^a I_a  \label{eq:cplxstr}
\end{equation}
for $s^1, s^2, s^3\in\R$. We easily read off
\begin{equation}
  J^2 = g_{ab} s^a s^b \stackrel{!}{=} -1 \quad \Leftrightarrow \quad
    (s^1)^2 + (s^2)^2 - (s^3)^2 = -1 .
\end{equation}
Obviously, the $\{s^a\}$ parametrize a two-sheeted hyperboloid~$H^2$.
We can map the upper half $H^2_+$ of $H^2$ onto the interior of the unit
disk in the $y$-plane by a stereographic projection
\begin{equation}
  s^1 := \frac{2y^1}{1-r^2}, \quad s^2 := \frac{2y^2}{1-r^2}, \quad
    s^3 := \frac{1+r^2}{1-r^2}, \quad r^2:=(y^1)^2+(y^2)^2 .
\end{equation}
Simultaneously, the lower half $H^2_-$ is mapped onto the exterior of
the unit disk. If we define $\l:=-(y^1+\ic y^2)$, both regions are
related by the map
\begin{equation}
  \s\colon\l\mapsto 1/\lb . \label{eq:sdef}
\end{equation}
Note that for $|\l|\neq 1$, $\s$ has no fixed points.
Recapitulating, we can state that the moduli space of complex
structures on $\R^{2,2}$ is $\C P^1 \backslash S^1$ (the $S^1$ being
given by $|\l|=1$).

\noindent
{\bf (Anti)holomorphic vector fields.}
A given complex structure~$J$ on~$\R^{2,2}$ as in~(\ref{eq:cplxstr})
has holomorphic and antiholomorphic eigenvectors with eigenvalues~%
$\ic$ and~$-\ic$; as a (local) basis for antiholomorphic vector fields,
we can choose
\begin{subequations} \label{eq:antihol}
\begin{align}
  \ov{W}_1 = \ov{W}_1^\mu \pa_\mu & = \frac{1}{2} (\pa_1+\ic\pa_2)
    -\frac{\l}{2}(\pa_3-\ic\pa_4) = \pa_{\zb^1} - \l\pa_{z^2} , \\
  \ov{W}_2 = \ov{W}_2^\mu \pa_\mu & = \frac{1}{2} (\pa_3+\ic\pa_4)
    -\frac{\l}{2}(\pa_1-\ic\pa_2) = \pa_{\zb^2} - \l\pa_{z^1} .
\end{align}
\end{subequations}
Their components, $\ov{W}_1^\mu=(\frac{1}{2},\frac{\ic}{2},-\frac{\l}{2},
\frac{\ic\l}{2})$ and $\ov{W}_2^\mu=(-\frac{\l}{2},\frac{\ic\l}{2},
\frac{1}{2},\frac{\ic}{2})$ satisfy $J_\mu{}^\nu \ov{W}_1^\mu =
-\ic \ov{W}_1^\nu$, $J_\mu{}^\nu \ov{W}_2^\mu = -\ic \ov{W}_2^\nu$,
and $\eta_{\mu\nu}\ov{W}_i^\mu\ov{W}_j^\nu=0$ for $i,j=1,2$,
respectively. The vector fields~(\ref{eq:antihol}) will become our Lax
operators subsequently (cf.~eqs.~(\ref{eq:gflax2})). The definitions of
$z^1$, $z^2$ coincide with those given in~(\ref{eq:cplx}). We can
introduce coordinates $\eta^1=z^1+\l \zb^2$ and $\eta^2=z^2+\l \zb^1$
(cf.~eq.~(\ref{eq:wprdef})) in the kernel of~(\ref{eq:antihol}).

\noindent
{\bf Almost complex structure on $\mathbf{H^2}$.}
If we introduce the standard complex structure $\eps$ on $H^2$,
\begin{equation}
  \eps^k_i \eps_k^j = -\de_i^j, \quad \eps_1^2=-\eps_2^1=1,
    \quad \eps_1^1=\eps_2^2=0
\end{equation}
for $i,j,k=1,2$, we can give explicit expressions for the (local)
antiholomorphic vector field on $H^2_+\subset\C P^1$:
\begin{equation} \label{eq:antihol2}
  \ov{W}_3=-\frac{1}{2}\left( \frac{\pa}{\pa y^1} + \ic\frac{\pa}
    {\pa y^2}\right) = \frac{\pa}{\pa\lb}.
\end{equation}

\noindent
{\bf Noncommutative description.}
In the noncommutative framework, one has to incorporate some modifications
to the above description. All functions now have to be multiplied by
a deformed product; alternatively, the Moyal-Weyl map may be used to
transform them into operators with the usual operator product. In this
interpretation, the space-time manifold $\R^{2,2}$ has to be replaced by
the Heisenberg algebra $\Acal$ generated by operators $\hat{x}^{\mu}$
subject to $[\hat{x}^{\mu},\hat{x}^{\nu}] = \ic \th^{\mu \nu}$. The
(Lie algebra of) inner derivations of $\Acal$ corresponds to the (Lie
algebra of) sections of the tangent bundle $T\R^{2,2}$. If we denote
by $\mathfrak{R}$ a four-dimensional representation of $SO(2,2)$, the
Lie algebra of inner derivations can be understood as a free
$\mathfrak{R}$-module. From these arguments it is clear that the
construction of the moduli space of complex structures on $\Acal$ can
be treated analogously to the commutative setup. Note that $H^2$
remains commutative.

Let us reconsider the noncommutative setup from a different point of view.
To this aim, we assume without loss of generality that $\th=\tht>0$.
Then, just as $z^1$ and $z^2$ are mapped to annihilation operators~%
(\ref{eq:annop}) under the Moyal-Weyl map, $\eta^1$ and $\eta^2$ are
mapped to new annihilation operators
\begin{equation}
  c_1:=(1-\l\lb)^{-1/2}\big( a_1+\l a_2^\dag \big) \quad\text{and}\quad
  c_2:=(1-\l\lb)^{-1/2}\big( a_2 + \l a_1^\dag \big), \label{eq:newannop}
\end{equation}
where $|\l|<1$. The $SO(2,2)$ rotation of the commutative discussion
above transforming old coordinates $z^i$ to new coordinates $\eta^i$
after transition to operators takes the form of a Bogoliubov
transformation~(\ref{eq:newannop}). In general, transformations $c_1=U
a_1 U^\dag$, $c_2=U a_2 U^\dag$ yield equivalent representations of the
Heisenberg algebra~$\Acal$ if $U$ is unitary. One can easily show that
this is the case for $|\l|\neq 1$ for~(\ref{eq:newannop}). Obviously,
the Bogoliubov transformations leaving a given representation invariant
are parametrized by the maximal pseudo-unitary subgroup $U(1,1)$ of
$SO(2,2)$ leading again to the same coset space as in the commutative
case.

\subsection{Real isotropic coordinates}
\noindent
Although $|\l|=1$ according to the preceding discussion will not
correspond to a complex structure on~$\R^{2,2}$, the vector fields~%
(\ref{eq:antihol}) in this case still span a null plane
in $\R^{2,2}$~\cite{Lechtenfeld:1999ik}. Using that now $\l=\lb^{-1}$,
one readily sees that complex conjugation maps $\ov{W}_1$ to a multiple
of $\ov{W}_2$ and vice versa, i.e., the isotropic plane is real. One is
free to choose a real basis for this plane, which is most easily
accomplished with the help of the map~(\ref{eq:trafo}) sending the
unit circle to the real axis in the $\z$-plane.

Real isotropic planes, being parametrized by $S^1=\{\l\in\C\big{|}
\,|\l|=1\}$, supplement the moduli space of complex isotropic two-%
planes (or complex structures) to $\C P^1\cong H^2\cup S^1$~%
\cite{Ivanova:2000zt, Lechtenfeld:1999ik}. So, $\C P^1$ can be
considered as the moduli space of all null two-planes (or extended
complex structures) in $\R^{2,2}$.

\subsection{Extended twistor space for $\R^{2,2}$} \label{sec:tw}
\noindent
{\bf Extended twistor space.}
In this section, Ward's theorem~\cite{Ward} on a one-to-one
correspondence between vector bundles~$E$ with self-dual connections
over euclidean~$\R^4$ and holomorphic bundles~$E'$ over the so-called
twistor space is rephrased for the case of $\R^{2,2}$. The twistor
space for $\R^{2,2}$ is the bundle~$\R^{2,2}\times H^2\to\R^{2,2}$ of
all constant complex structures on $\R^{2,2}$. It can be endowed with
the direct sum $\Jcal$ of the complex structures $J$ and $\eps$. The
vector fields~(\ref{eq:antihol}) and~(\ref{eq:antihol2}) for $|\l|
\neq 1$ are the $\Jcal$-antiholomorphic vector fields on~$\R^{2,2}
\times H^2$ with respect to this complex structure. Admitting
$|\l|=1$ in~(\ref{eq:antihol}) and~(\ref{eq:antihol2}), we can extend
these vector fields naturally to $\Zcal:=\R^{2,2}\times\C P^1$.

{\bf Vector bundle over $\mathbf{\Zcal}$.}
Now, we can use the canonical projection $\pi\colon\Zcal\to\R^{2,2}$ to
lift the vector bundle~$E:=P\times_{GL(N,\C)} \C^N$ to a bundle~%
$\pi^* E$ over the extended twistor space~$\Zcal$. By construction, the
connection on~$\pi^* E$ is flat along the fibers $\C P^1$ of~$\Zcal$, so
that the lifted connection~$\pi^* A$ on~$\pi^* E$ can be chosen to have
only components along $\R^{2,2}$, $\pi^* A = A_\mu dx^\mu$. Thus, the
lift takes the covariant derivative $D_\mu=\pa_\mu + A_\mu$ on $E$ to
\begin{equation}
  \pi^* D = dx^\mu D_\mu + dy^i\frac{\pa}{\pa y^i} \label{eq:liftconn}
\end{equation}
on $\pi^* E$. Now, the $\Jcal$-antiholomorphic components of~%
(\ref{eq:liftconn}) are the $(0,1)$ components of $\pi^* D$ along
the antiholomorphic vector fields $\ov{W}_i$ on~$\Zcal$:
\begin{subequations} \label{eq:laxops}
\begin{eqnarray}
  D_1^{(0,1)} \,\equiv\, \ov{W}_1^\mu D_\mu & = & \ov{W}_1 + \frac{1}{2}
    (A_1+\ic A_2)-\frac{\l}{2} (A_3-\ic A_4) , \\
  D_2^{(0,1)} \,\equiv\, \ov{W}_2^\mu D_\mu & = & \ov{W}_1 + \frac{1}{2}
    (A_3+\ic A_4)-\frac{\l}{2} (A_1-\ic A_2) , \\
  D_3^{(0,1)} \,\equiv\, \ov{W}_3^i \pa_{y^i} & = & \ov{W}_3 .
\end{eqnarray}
\end{subequations}

\noindent
{\bf Holomorphic sections.}
Local sections~$\varphi$ of the complex vector bundle~$\pi^* E$ are
holomorphic if
\begin{subequations} \label{eq:holsec}
\begin{align}
  D_1^{(0,1)} \varphi = 0 , \\
  D_2^{(0,1)} \varphi = 0 , \\
  D_3^{(0,1)} \varphi = 0 . \label{eq:thrdlax}
\end{align}
\end{subequations}
We can also view this as the local form of meromorphic sections of
$E':=\pi^* E$ in a given trivialization of the bundle. One can combine
$N$ such sections (as columns) into an $N\times N$ matrix to obtain the
matrix-valued function~$\psi$ used in~(\ref{eq:laxpair2}). Using~%
(\ref{eq:laxops}), a comparison with~(\ref{eq:laxpair2}) shows
that after solving~(\ref{eq:thrdlax}) these are exactly the linear
equations (Lax pair) for self-dual Yang-Mills theory. In this framework,
the self-duality equations~(\ref{eq:sdcomplex}) emerge as the condition
that eqs.~(\ref{eq:holsec}) are compatible, i.e., the $(0,2)$~components
of the curvature of~$E'$ vanish.

\subsection{Reality condition}
\noindent
So far, we have been working with a complex vector bundle associated
to a $GL(N,\C)$-principal bundle $P$ to describe $U(N)$ self-dual
Yang-Mills theory. Therefore, we have to implement a reality condition
on our gauge fields, i.e., impose the additional constraint $A_\mu^\dagger=
-A_\mu$.

Let us now scrutinize the action of hermitean conjugation on the
linear equations~(\ref{eq:laxpair2}). Eq.~(\ref{eq:laxpair2a}) is
equivalent to
\begin{equation}
  (\pa_{\zb^1}-\l\pa_{z^2})\psi^{-1}(\l) = \psi^{-1}(\l)
    (A_{\zb^1}-\l A_{z^2}) ,
\end{equation}
where we suppress the additional dependence of $\psi$ of the space-time
coordinates~$z^i, \zb^i$. Since we demand this to hold for all~$\l$, we
can as well first apply~$\s$ from~(\ref{eq:sdef}) and then take the
hermitean conjugate,
\begin{equation}
  (\pa_{\zb^2}-\l\pa_{z^1})\left[\psi^{-1}(\lb^{-1})\right]^\dagger =
    -(A_{\zb^2}-\l A_{z^1})\left[\psi^{-1}(\lb^{-1})\right]^\dagger .
\end{equation}
This coincides with~(\ref{eq:laxpair2b}) if we choose $\psi(\l) =
\left[ \psi(\lb^{-1})^\dagger \right]^{-1}$, i.e., eq.~(\ref{eq:reality2}).
With these restrictions, the $gl(N,\C)$-curvature~$F$ naturally descends
to a $u(N)$-valued curvature.

In the noncommutative case, the vector bundle $E$ has to be
replaced by a free module over $\Acal$. Accordingly, $D=d+A$ is chosen
to be a connection on the module $E$ \cite{Konechny:2001wz}. It is
understood that the above discussion can be applied analogously,
taking into account that multiplication of $A$ and $\psi$ becomes
noncommutative.

\section{Abelian pseudo-instantons}\label{sec:energy}
This appendix concludes our considerations with the discussion of a
special class of abelian, i.e., $U(1)$ solutions with finite action
(in contrast to their commutative counterparts).\footnote{Noncommutative
instantons in euclidean space were introduced in~\cite{Nekrasov:1998ss}.}
We work in the operator formalism. Let us introduce ``shifted'' operators
acting on the two-oscillator Fock space $\Hcal$:
\begin{equation}
  X_\mu := A_\mu + \ic(\th^{-1})_{\mu\nu}x^\nu ,
\end{equation}
where $(\th^{-1})_{\mu\s}\th^{\s\nu}=\de_\mu^\nu$. The operator-valued
field strength $F_{\mu\nu}$ can be expressed in terms of the shifted
operators $X_\mu$ as
\begin{equation}
  F_{\mu\nu}= [X_\mu, X_\nu] - \ic(\th^{-1})_{\mu\nu}.
\end{equation}
The incarnation of the ncYM equations in this context is
\begin{equation}
  [X^\mu, [X_\mu,X_\nu]] = 0.
\end{equation}
They are, of course, automatically satisfied by $X_\mu$ subject to the
ncSDYM equations (cf.~\cite{Franco-Sollova:2002nn})
\begin{equation}\label{eq:opncsdym}
  [X_\mu,X_\nu] = \frac{1}{2}\eps_{\mu\nu\r\s} [X^\r,X^\s] + \ic
    (\th_{\mu\nu} - \frac{1}{2}\eps_{\mu\nu\r\s}\th^{\r\s}).
\end{equation}
Observe that the last term of the ncSDYM equations vanishes for self-%
dual $\th^{\mu\nu}$, i.e., $\th=\tht$. Switching to complex coordinates%
\footnote{We denote $[z^i,\zb^j]=\ic\th^{i\jb}$ and $(\th^{-1})_{\ib k}
\th^{k\jb}=\de_\ib^\jb$.} and assuming self-dual $\th^{\mu\nu}$ from now
on, eqs.~(\ref{eq:opncsdym}) become
\begin{subequations}\label{eq:opcplxsdym}
\begin{align}
  [X_{z^1},X_{z^2}] = [X_{\zb^1},X_{\zb^2}] & = 0, \\
  [X_{z^1},X_{\zb^1}]-[X_{z^2},X_{\zb^2}] & = 0,
\end{align}
\end{subequations}
where $X_{z^i}:= A_{z^i} + \ic (\th^{-1})_{i\jb} \zb^j$ and
$X_{\zb^i}:= A_{\zb^i} + \ic (\th^{-1})_{\ib j} z^j$, $i\in \{1,2\}$.
It is easily checked that
\begin{subequations}
\begin{align}
  X^0_{z^1} = \ic (\th^{-1})_{1 \bar{1}} \zb^1,
  & \qquad X^0_{\zb^1} = \ic (\th^{-1})_{\bar{1} 1}z^1, \\
  X^0_{z^2} = \ic (\th^{-1})_{2 \bar{2}}\zb^2,
  & \qquad X^0_{\zb^2} = \ic (\th^{-1})_{\bar{2} 2}z^2,
\end{align}
\end{subequations}
i.e., $A_{z^i}=A_{\zb^i}=0$ yields a (trivial) solution of~%
(\ref{eq:opcplxsdym}).

New solutions $X^1$ may be obtained by shift operator ``dressing'' of
the solutions $X^0$, namely
\begin{subequations}
\begin{align}
  X^1_{z^1} = S X^0_{z^1} S^\dag, & \quad X^1_{\zb^1} = S X^0_{\zb^1}
    S^\dag, \\
  X^1_{z^2} = S X^0_{z^2} S^\dag, & \quad X^1_{\zb^2} = S X^0_{\zb^2}
    S^\dag .
\end{align}
\end{subequations}
In these expressions, $S$ and $S^{\dag}$ are shift operators acting on
the two-oscillator Fock space $\Hcal$ according to
\begin{equation}
  S^\dag S = 1, \quad SS^\dag = 1 - P_0, \quad P_0 S = S^\dag P_0 = 0.
\end{equation}
Apparently, the representation of $S$ on $\Hcal$ is not unique (see,
e.g., \cite{Aganagic:2000mh, Lechtenfeld:2001ie} for various explicit
forms of $S$ and $S^{\dag}$). Here, $P_0$ denotes
the projector onto the ground state $|0,0\rangle$ of the Fock space
$\Hcal$:
\begin{equation}
  P_0 = |0,0\> \< 0,0|.
\end{equation}

The field strength for such configurations turns out to be of the form
\begin{equation}\label{eq:AGMS}
  F_{z^i\zb^i} = [X^1_{z^i},X^1_{\zb^i}] -\ic (\th^{-1})_{i\ib}
   = - \ic (\th^{-1})_{i\ib} P_0 = -\frac{1}{2\th} P_0,
    \quad i\in\{1,2\}.
\end{equation}
This coincides with the solution first presented in~%
\cite{Aganagic:2000mh} for the euclidean case, namely on $\R^4$.
The action for this type of solution is known to be finite; this is
also the case here:
\begin{equation}
  S_1 = -\frac{1}{2g^2_{YM}}(2\pi\th)^2
    \text{Tr}_{\Hcal}\; \Fh_{z^i \zb^j} \Fh^{z^i \zb^j}
    = \frac{4\pi^2}{g^2_{YM}}.
\end{equation}
In the context of D-branes, solutions of type~(\ref{eq:AGMS}) have been
interpreted as a D-brane of codimension four sitting at the origin of a
space-time filling D-brane~\cite{Aganagic:2000mh}. This may be transferred
to our case.
\end{appendix}

\end{document}